\RequirePackage{fix-cm}
\documentclass[smallextended]{svjour3}       
\smartqed  
\usepackage{graphicx}

\usepackage{amsmath} 
\usepackage[usenames,dvipsnames]{color}
\usepackage{url}
\definecolor{linkcolor}{rgb}{0,0,0.6} 
\usepackage[colorlinks=true, pdfstartview=FitV, linkcolor= linkcolor, citecolor= linkcolor, urlcolor= linkcolor, hyperindex=true,hyperfigures=true]{hyperref}
\usepackage{mathptmx}      
\newcommand{\e}{{\mathrm{e}}}

\newcommand{\tot}{{\mathrm{tot}}}
\newcommand{\sys}{{\mathrm{sys}}}
\newcommand{\med}{{\mathrm{med}}}

\newcommand{\SC}{{\kappa}}
\newcommand{\fric}{\widehat{\gamma}}

\newcommand{\CovM}{{\mathcal{S}}}

\begin{document}

\title{Explicit solution of the Generalised Langevin equation
}


\author{Ivan Di Terlizzi \and Felix Ritort \and
        Marco Baiesi 
}


\institute{Ivan Di Terlizzi and Marco Baiesi \at
           Dipartimento di Fisica e Astronomia,
           Universit\`a di Padova, Via Marzolo 8, 35131, Padova, Italy
           \\
           \at
           INFN, Sezione di Padova, Via Marzolo 8, 35131, Padova, Italy
           \and
           Felix Ritort
           \at Condensed Matter Physics Department, University of Barcelona, C/Marti i Franques s/n, 08028 Barcelona, Spain
}

\date{Received: date / Accepted: date}

\maketitle
\begin{abstract}
Generating an initial condition for a Langevin equation with memory is a non trivial issue. We introduce a generalisation of the Laplace transform as a useful tool for solving this problem, in which a limit procedure may send the extension of memory effects to arbitrary times in the past. This method allows us to compute average position, work, their variances and the entropy production rate of a particle dragged in a complex fluid by an harmonic potential, which could represent the effect of moving optical tweezers. For initial conditions in equilibrium we generalise the results by van Zon and Cohen, finding the variance of the work for generic protocols of the trap. In addition, we study a particle dragged for a long time captured in an optical trap with constant velocity in a steady state. Our formulas open the door to thermodynamic uncertainty relations in systems with memory.
\end{abstract}

\section{Introduction}

The driven diffusion process of a colloidal particle or bead immersed in a fluid has become a paradigm of nonequilibrium physics~\cite{maes03,VanZon_cohen_Work,bai06,ron07,sek10,sei12,cil19}. Fluctuations play a prominent role for this mesoscopic system due to the multitude of random hits on the particle by the molecules of the surrounding fluid. If these molecules are tinier and faster than the colloidal particle, a net separation of timescales between fast and slow degrees of freedom occurs and the colloidal particle undergoes Markovian dynamics. In this case, the motion of the particle can be equivalently described by using the Langevin equation, path integrals and the Fokker-Plank equation~\cite{ris89}. Historically, the Langevin approach came first and arguably remains the most intuitive. In fact, for a one dimensional system, by incorporating the effects of the fluid in Newton's second law one may write a Langevin equation of motion for the position $x(t)$ of a particle of mass $m$ as a second order stochastic differential equation,
\begin{equation}\label{LE3}
  m \ddot x (t) =  - \gamma_{0} \dot x (t) + \mathcal{F}(x,t)+ \xi (t) \,.
\end{equation}
The random force is generated by a Gaussian white noise $\xi(t)$, with average $\langle \xi(t) \rangle=0$ and correlation $\langle \xi(t') \xi(t'') \rangle = 2  \gamma_{0}k_B T \delta(t'-t'')$. The prefactor of the delta function ensures thermodynamic consistency according to the (second) fluctuation-dissipation theorem~\cite{kub66}, linking the drag coefficient $\gamma_{0}$ of the dissipative term $- \gamma_{0} \dot x$ to the strength of the noisy term. As a deterministic force not due to the fluid we focus on the case $\mathcal{F}(x,t)= - \partial_{x}U(x,t)$ with a time-dependent potential energy $U(x,t)$.

If the particle is immersed in a solution containing for example long and complex polymers \cite{FractionalProteins1,FractionalProteins2}, the above-mentioned separation of time scales is no longer possible and memory effects occur.
One may then consider a generalised Langevin equation (GLE) with constant diffusion coefficient, whose formal derivation can be found in \cite{zwanzig2001nonequilibrium,weiss2012quantum,MoriGLE}. For $t\ge 0$ this equation reads
\begin{equation}\label{GLE1}
    m \ddot x (t) =  - \int_{t_m}^{t}\mathrm{d}t'\Gamma(t-t') \dot x (t')  - \partial_{x}U(x,t) + \eta (t) \, ,
\end{equation}
where $\Gamma(t)$ is the memory kernel, $t_m \le 0$ is the time to which the memory effects extend and $\eta(t)$ is a coloured Gaussian noise obeying  $\langle \eta(t) \rangle=0$. The above equation could also describe the motion of a particle under the effect of hydrodynamic backflow \cite{BackflowGLE}. The  fluctuation-dissipation relation~\cite{kub66} is still valid in the more general form
$\langle \eta(t') \eta(t'') \rangle =  k_{B}T \Gamma(|t'-t''|)$:
thermodynamic equilibrium is present in the medium if its two effects (dissipation and noise) are proportional at all times. Note that a Markovian memory kernel $\Gamma^{\text{Markov}}(t)= 2 \gamma_{0} \delta(t)$ would lead to the usual fluctuation-dissipation theorem (the consistency, instead, of equation \eqref{GLE1} with the usual Langevin equation for $\Gamma^{\text{Markov}}(t)= 2 \gamma_{0} \delta(t)$, is guaranteed by the Stratonovich convention for the integrals of delta functions, i.e. $\int_{t_m}^{t}\mathrm{d}t'f(t')\delta(t'-t) = f(t)/2$)).

The aim of this paper is to solve the GLE with a parabolic confinement potential $U(x,t) = \frac \SC 2 (x-\lambda(t))^2$, obtained for instance by using optical tweezers centred on a moving coordinate $\lambda(t)$,
\begin{equation}\label{GLE2}
    m \ddot x (t) =  - \int_{t_m}^{t}\mathrm{d}t'\Gamma(t-t') \dot x (t')  - \SC [x(t)-\lambda(t)] + \eta (t) \,.
\end{equation}
The non-dynamical case was already discussed for example in \cite{LisyGLE}. Moreover, we will restrict ourselves to the case of a non-divergent time dependent effective friction coefficient $\fric(t)$, i.e. such that $\fric = \displaystyle \lim_{t \to \infty} \fric(t) = \displaystyle \lim_{t \to \infty} \int_{0}^{t}\mathrm{d}t'\Gamma(t') < \infty$, which is a sensible physical requirement~\cite{ViscGLE_Goy,Molina_Garcia_2018}.

One of the first analytical solutions for the GLE with $\SC=0$ and no external force can be found in \cite{Fox_GLE}. It is obtained through the use of Laplace transforms and it is expressed in terms of the velocity susceptibility $\chi_{v}(t)$, a key quantity discussed in the next sections. In this paper we  obtain a more general solution in terms of the susceptibility and its integrals.
This enables us to calculate averages and variances of relevant quantities such as position, thermodynamic work and entropy production, with a dynamics starting from different initial conditions. Some of these results are already known in the literature, especially for equilibrium initial conditions, see  for example~\cite{WorkFT_NonMarkov}. However, imposing a nonequilibrium steady state as initial condition is not trivial for the GLE, due to its memory. A scheme for achieving an initial condition with memory requires extending it far into the past. To this end, we introduce a modified version of Laplace transforms with arbitrary initial time $t_m$, which is then shifted back to minus infinity by taking an appropriate limit. The explicit dependence of the solution on $t_m$ along with the well-defined limits of susceptibilities will make the procedure straightforward.

The following section introduces the technical details of the modified Laplace transform. In Sec.~\ref{sec:GLEsol} we discuss the solution of the GLE and in Sec.~\ref{sec:thermo} we show how to use the solution for computing relevant thermodynamic quantities. We show that the entropy production rate can be expressed in terms of a retarded velocity, which is equal to the usual velocity of the particle in the Markovian case, see \eqref{ret_vel}. In section \ref{sec:over} we briefly discuss the overdamped case, corresponding to $m=0$. Moreover, in Sec.~\ref{sec:appl}, we apply the obtained results to the dynamics starting from equilibrium and to the case where initial conditions are taken in the infinite past, i.e. $t_m \to - \infty$, which can be seen as a generalised stationary state, in the sense that memory of initial conditions is lost. For the latter case we manage to show that the variance of the thermodynamic work is equal to that of a system prepared in equilibrium initial conditions for every driving protocol $\lambda(t)$ (see equation\eqref{varWeqstaz}), thus generalising the results by van Zon and Cohen~\cite{VanZon_cohen_Work}. Finally we consider the special case of a linear dragging protocol $\lambda(t)=vt$ with $t_m \to - \infty$, also discussed in \cite{berner2018oscillating}, which can be considered as a steady state in the usual sense. For this scenario we show that quantities such as average position, velocity, work and entropy production rate have the same structure as for Markov dynamics. The variances, however, are different.

\section{Modified Laplace transform}

A standard way of dealing with the linear GLE uses Laplace transforms. This technique is particularly useful when dealing with an initial condition at finite times, for instance when the system starts from equilibrium at time $t=0$. If the initial time is rather taken infinitely back  in the past, traditional Laplace transforms are no longer suitable to find a solution for the GLE. However, it is well known that, for Markovian dynamics, non-equilibrium steady states can be obtained from this limit. Hence, we would find it useful to have a framework in which Laplace transforms are available and steady states may be considered.

Our way to tackle this problem is to introduce a modified Laplace transform with an arbitrary initial time $t_m\le 0$ that acts on a given function $g(t)$ as follows
\begin{equation} \label{Laplace1}
  \hat{g}^{t_{m}}(k) = \mathcal{L}^{t_{m}}_{k} [g(t)] = \int_{t_{m}}^{\infty}\mathrm{d}t ~ \e^{-kt} g(t) \, .
\end{equation}
The standard Laplace transform of course is recovered for $t_{m}\nearrow 0$.

The aim is to solve the GLE finding the explicit dependence of the solution on $t_m$ and then, if interested in steady states, eventually take the limit $t_m \to -\infty$.
For our purposes, we just need to know the effect of such modified transform on first and second derivatives of a function. They can be readily expressed as
\begin{equation}\label{Lap1}
  \begin{split}
    & \hspace{1.2cm} \mathcal{L}^{t_{m}}_{k} [\dot{g}(t)] = k \hat{g}^{t_{m}}(k)-g_{t_{m}}\e^{-k t_{m}} \, ,  \\
    & \mathcal{L}^{t_{m}}_{k} [\ddot{g}(t)] = k^{2} \hat{g}^{t_{m}}(k)-k g_{t_{m}}\e^{-k t_{m}}- \dot{g}_{t_{m}}\e^{-k t_{m}}\,.
  \end{split}
\end{equation}  
Note that $\hat{g}^{t_{m}}(k)$ stands for the modified Laplace transform of the function $g(t)$ while $g_{t_m} \equiv g(t_m)$ is the function calculated at time $t_m$.

Furthermore, it is not hard to show that the action of the modified Laplace transform on integrals is equal to the action of the standard transform, namely
\begin{equation}
    \mathcal{L}^{t_{m}}_{k} \left[\int_{t_m}^{t}\mathrm{d}t'~g(t')\right] = \frac{\hat{g}^{t_{m}}(k)}{k} \, .
\end{equation}
 We also need to know the effect of such transform on the convolution of a causal function $\mathcal{G}(t)$, i.e such that $\mathcal{G}(t<0)=0$ (like the memory kernel $\Gamma(t)$ in our case), with an arbitrary $g(t)$
\begin{equation} \label{conv1}
  \mathcal{L}^{t_{m}}_{k} \left[\int_{t_m}^{t}\mathrm{d}t'\mathcal{G}(t-t')g(t')\right] = \int_{t_{m}}^{\infty}\mathrm{d}t \int_{t_m}^{t}\mathrm{d}t'\e^{-kt}\mathcal{G}(t-t')g(t')\,.
\end{equation}
First, to compute an explicit version of this equation, we note that 
\begin{equation} \label{Domain}
  \int_{t_{m}}^{\infty} \mathrm{d}t \int_{t_m}^{t}\mathrm{d}t' = \int_{t_{m}}^{\infty}\mathrm{d}t' \int_{t'}^{\infty}\mathrm{d}t
\end{equation}
i.e.~these integrals define the same region of integration so that \eqref{conv1} becomes
\begin{equation}
\begin{split}
  &\hspace{1.5cm}\mathcal{L}^{t_{m}}_{k} \left[\int_{t_m}^{t}\mathrm{d}t'\mathcal{G}(t-t')g(t')\right] = \int_{t_{m}}^{\infty}\mathrm{d}t' \int_{t'}^{\infty}\mathrm{d}t~\e^{-kt}\mathcal{G}(t-t')g(t')= \\
  & \stackrel{u=t-t'}{=} \int_{t_{m}}^{\infty}\mathrm{d}t' \int_{0}^{\infty}\mathrm{d}u~\e^{-ku}\e^{-kt'}\Gamma(u)g(t')= \int_{t_{m}}^{\infty}\mathrm{d}t'\e^{-kt'}g(t') \int_{0}^{\infty}\mathrm{d}u~\e^{-ku}\mathcal{G}(u) = \\
  & \hspace{3.5cm} = \mathcal{L}^{t_{m}}_{k} \left[g(t)\right]\mathcal{L}_{k} \left[\mathcal{G}(t)\right]= \hat{g}^{t_m}(k) \hat{\mathcal{G}}(k)
\end{split}
\end{equation}
which is a generalisation of the convolution theorem. It states that the modified Laplace transform of the convolution of a causal function $\mathcal{G}(t)$ with an arbitrary function $g(t)$ is equal to the product of the standard Laplace transform of the causal function, i.e. $\hat{\mathcal{G}}(k)$, and the modified Laplace transform of $g(t)$, that is $\hat{g}^{t_m}(k)$. 

We conclude this section by remarking that, of course, the modified Laplace transform of a causal function is equal to the standard Laplace transform of that function.

\section{GLE solution}\label{sec:GLEsol}

\begin{figure}[tb]
\includegraphics[width=0.78\textwidth, angle=0]{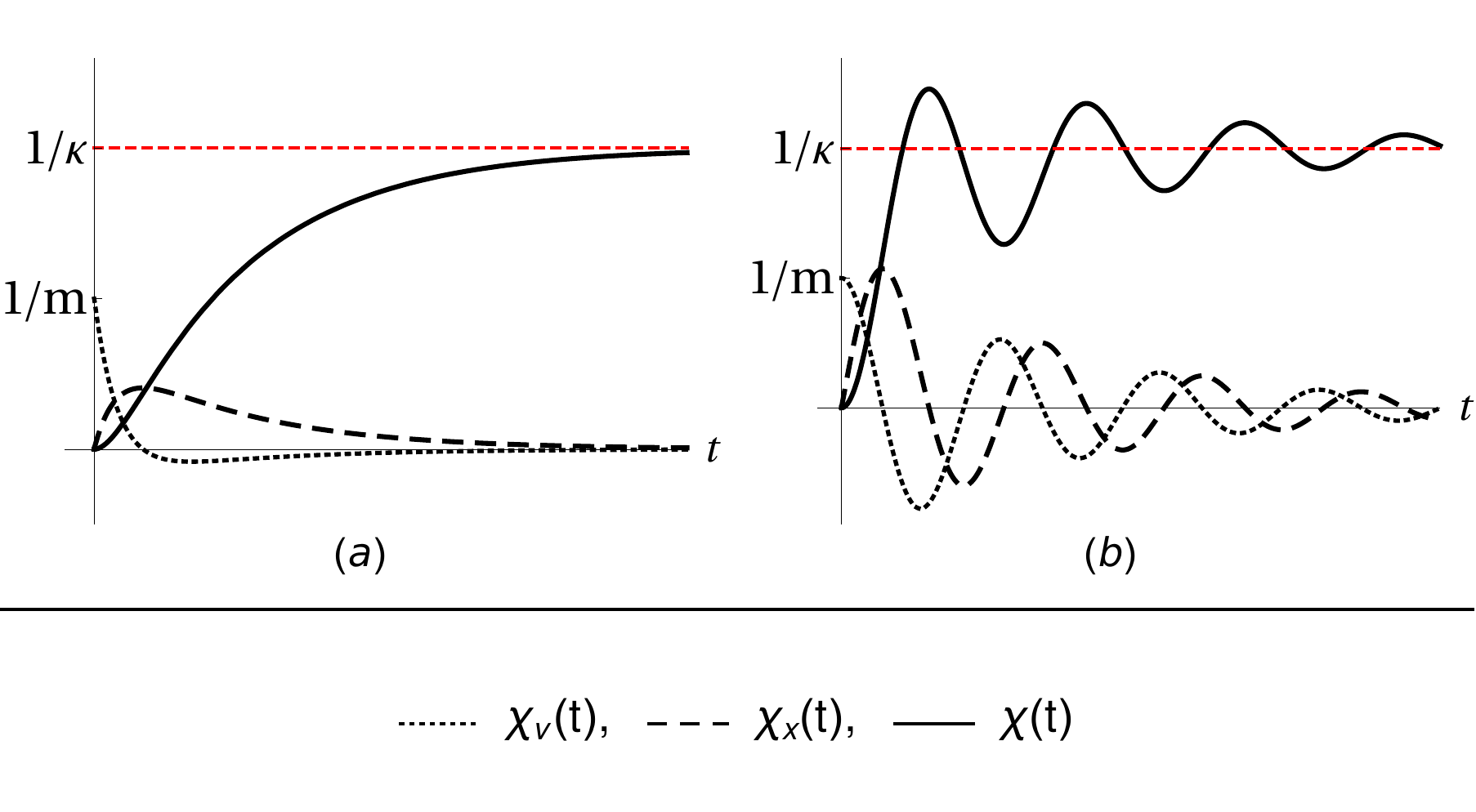}
\caption{Underdamped ($m\neq0$) susceptibilities for (a) Markovian memory kernel $\Gamma^{\text{Markov}}(t)= 2 \gamma_0 \delta(t)$ and (b) for non-Markovian memory kernel of the form $\Gamma^{\text{exp}}(t)= (\gamma/\tau)\exp[-t/\tau]$. In both cases we see that $\displaystyle \lim_{t \to 0} \chi_{v}(t)=1/m$, $\displaystyle \lim_{t \to \infty} \chi_{v}(t)=0$, $\displaystyle \lim_{t \to 0} \chi_{x}(t)=0$, $\displaystyle \lim_{t \to \infty} \chi_{x}(t)=0$, $\displaystyle \lim_{t \to 0} \chi(t)=0$ and $\displaystyle \lim_{t \to \infty} \chi(t)=1/\SC$. In this underdamped case, all the mentioned limits remain valid for all memory kernels, see Appendix \ref{sec:limits}}.
\label{Fig1}
\end{figure}

By applying the modified Laplace transform \eqref{Laplace1} to the GLE \eqref{GLE2} and by using the results obtained above
\begin{equation}
  \mathcal{L}^{t_{m}}_{k} \left[ m\ddot x (t) \right] = \mathcal{L}^{t_{m}}_{k} \left[- \int_{t_{m}}^{t}\mathrm{d}t'\Gamma(t-t') \dot x(t') - \SC (x (t)-\lambda(t)) + \eta (t)\right]
\end{equation}
we get
\begin{equation} \label{sol1} 
\begin{split}
  &\hspace{3.5cm} m \left(k^{2} \hat{x}^{t_m}(k) - k x_{t_{m}} \e^{-k t_{m}} - v_{t_{m}} \e^{-k t_{m}} \right) =\\
  & \hspace{2cm}- \hat{\Gamma}(k) \left[ k \hat{x}^{t_{m}}(k)-x_{t_{m}}\e^{-k t_{m}} \right] - \SC \hat{x}^{t_m}(k) + \SC \hat{\lambda}^{t_m}(k) + \hat{\eta}^{t_m}(k) \,.
\end{split}
\end{equation}
Furthermore, with a bit of algebra we can isolate the position $x$ from the other quantities obtaining
\begin{equation}
   \hat{x}^{t_{m}}(k) = x_{t_{m}} \frac{\e^{-k t_{m}}}{k}(1-\SC \hat{\chi}_{x}(k))+m v_{t_{m}}\e^{-k t_{m}}\hat{\chi}_{x}(k) + (\SC \hat{\lambda}^{t_{m}}(k) + \hat{\eta}^{t_{m}}(k))\hat{\chi}_{x}(k) \, ,
\end{equation}
where we introduced the ``position susceptibility''  $\chi_x (t)$, a key quantity of this paper, defined via its Laplace transform 
\begin{equation} \label{chix}
\hat{\chi}_{x}(k) = [m k^{2}+k\hat{\Gamma}(k)+\SC]^{-1} \, .
\end{equation}
In the following we will also use its integral $\chi(t)$ and its derivative $\chi_{v}(t)$ (``velocity susceptibility'')
\begin{align} \label{chi}
  \chi(t) &\equiv \int_{0}^{t} \mathrm{d}t' \chi_{x}(t') \, , \\
  \chi_{v}(t) &\equiv \partial_{t} \chi_{x}(t) \, .
\end{align}
In Appendix~\ref{sec:limits} we discuss the limits of these susceptibilities for $t\to 0$ and $t\to\infty$.
Two examples are shown in figure~\ref{Fig1}.

We stress that all the susceptibilities are of course causal functions.

By defining the inverse of the modified Laplace transform through the usual Bromwich integral
\begin{equation}
    g(t) = \frac{1}{2 \pi i} \int_{\alpha -i \infty}^{\alpha +  i \infty} \mathrm{d}k \hspace{0.1cm} \e^{kt} \hspace{0.1cm} \hat{g}^{t_m} (s)\, ,
\end{equation}
where $\alpha$ is such that the chosen vertical contour in the complex plane has all the singularities of $g(s)$ on its left, we see that  $\mathcal{L}^{-1,t_{m}}_{k} \left[ \e^{-k t_{m}} \right]= 2 \delta(t-t_m)$ (the  factor $2$ is needed for consistency) and $\mathcal{L}^{-1,t_{m}}_{k} \left[ \frac{\e^{-k t_{m}}}{k} \right]=\theta (t-t_m)$, where $\theta(t)$ is the Heaviside step function. Transforming back equation \eqref{sol1} to real time we obtain, for  $t > 0 \geq t_m$,
\begin{equation}\label{solution}
\begin{split}
   x(t) =& x_{t_m} \left( \theta(t-t_m) - \SC \int_{t_m}^{t}\mathrm{d}t'\chi_{x} (t-t') \theta(t'-t_m) \right) +\\
  & + \textcolor{red}{2} m v_{t_m}\int_{t_m}^{t}\mathrm{d}t'\chi_{x} (t-t') \delta(t'-t_m) + \int_{t_m}^{t}\mathrm{d}t'\chi_{x} (t-t') \left[\SC \lambda(t')+\eta(t') \right] =\\
  =& x_{t_m} \left( 1 - \SC \chi (t-t_m) \right) + m v_{t_m}\chi_{x} (t-t_m) + \int_{t_m}^{t}\mathrm{d}t'\chi_{x} (t-t') \left[\SC \lambda(t')+\eta(t') \right] \, ,
\end{split}
\end{equation}
that is the solution to the generalised Langevin equation. The velocity can be readily obtained by simply taking its time derivative:
\begin{equation}\label{solution_velocity}
\begin{split}
   v(t) = -\SC   x_{t_m}  \chi_x (t-t_{m})+ m v_{t_{m}} \chi_{v}(t-t_m) + \int_{t_m}^{t}\mathrm{d}t'\chi_{v} (t-t') \left[\SC \lambda(t')+\eta(t') \right]\\
\end{split}
\end{equation}
where we used that for underdamped dynamics $\chi_x(0) = 0$, see Appendix \ref{sec:limits}. Taking the averages of the above expressions and using that $\langle \eta(t) \rangle = 0$, we get
\begin{equation} \label{avgx}
\begin{split}
  & \langle x \rangle_{t_m,t} = \langle x_{t_m} \rangle(1-\SC \chi (t-t_{m}))+ m\langle v_{t_{m}}\rangle \chi_{x}(t-t_m) + \SC \int^{t}_{t_{m}}\mathrm{d}t' \chi_{x} (t-t') \lambda(t')  \\
\end{split}
\end{equation}
\begin{equation} \label{avgv}
\begin{split}
   \langle v \rangle_{t_m,t} = -\SC \langle  x_{t_m} \rangle \chi_x (t-t_{m})+ m\langle v_{t_{m}}\rangle \chi_{v}(t-t_m) + \SC \int^{t}_{t_{m}}\mathrm{d}t' \chi_{v} (t-t') \lambda(t') \, ,
\end{split}
\end{equation}
with the notation $\langle \cdot \rangle_{t_m,t}$ meaning that initial conditions are taken at time $t_m$ while the observation time is taken at time $t$.

\subsection{Variance of the position and correlations}

Another important quantity we are interested in is the variance of the position at time $t$. Given that the system started at time $t_m$ with position $x_{t_m}$ and velocity $v_{t_m}$, we have that 
\begin{equation} \label{Var1}
  \langle \Delta^{2} x \rangle_{t_m,t} = \langle \left(x (t) - \langle x \rangle_{t_m,t} \right)^{2} \rangle_{t_m,t} \, .
\end{equation}
Using the previously obtained expression for the position \eqref{solution} and defining
\begin{equation} 
\phi(t) = \int^{t}_{t_{m}} \chi_{x} (t-t') \eta (t') \mathrm{d}t'\,,
\end{equation}
we find that \eqref{Var1} becomes
\begin{equation} \label{Var2}
\begin{split}
  \langle \Delta^{2} x \rangle_{t_m,t} =& \langle \phi^2 (t) \rangle + \langle \Delta^{2}x_{t_m}\rangle (1-\SC \chi (t-t_{m}))^{2}+ m^{2}\langle \Delta^{2} v_{t_{m}}\rangle \chi^{2}_{x}(t-t_m)\\
  & + 2m \textrm{Cov}(x_{t_m},v_{t_m})\chi_{x}(t-t_m)(1-\SC \chi (t-t_{m})) \, .
\end{split} 
\end{equation}
Focusing on the the first term on the right hand side, we further define the following quantity (also for future convenience):
\begin{equation}\label{Var3}
\begin{split}
  \mathcal{C} (t',t'')= \langle \phi (t')\phi (t'') \rangle = \int_{t_m}^{t'}\mathrm{d}s'\int_{t_m}^{t''}\mathrm{d}s'' \chi_{x} (t'-s') \chi_{x} \left( t''-s''\right) \langle \eta(s') \eta \left( s'' \right) \rangle \, ,
  \end{split}
\end{equation}
which in Appendix B we show to be equal to
\begin{equation}\label{C(t',t'')}
\begin{split}
  \mathcal{C} (t',t'') =& k_{B}T \Big[\chi (t'-t_m)+\chi(t''-t_m)-\theta(t'-t'')\chi(t'-t'') +\\
  &-\theta(t''-t')\chi(t''-t')- \SC\chi (t'-t_m)\chi(t''-t_m) - m\chi_{x} (t'-t_m)\chi_{x}(t''-t_m) \Big] \,.
  \end{split}
\end{equation}
The variance of the position can be obtained by evaluating this quantity at equal times (i.e.~$t=t'=t''$) and then by plugging it into equation \eqref{Var2}. From its definition \eqref{chi} one immediately sees that $\chi(0)=0$, so that
\begin{equation}
  \langle \phi^{2}(t)\rangle = \mathcal{C} (t,t) = k_{B}T \Big[2\chi (t-t_m) - \SC\chi^{2} (t-t_m) - m\chi_{x}^{2} (t-t_m) \Big] \,.
\end{equation}
Finally, by using \eqref{Var2}, we obtain an expression for the variance of the position from arbitrary initial conditions
\begin{equation} \label{varx}
\begin{split}
  \langle \Delta^{2} x \rangle_{t_m,t} =& k_{B} T \Big[ 2 \chi(t-t_m) - m \chi^{2}_{x}(t-t_m) - \SC \chi^{2}(t-t_{m}) \Big] +\\
  &+ \langle \Delta^{2}x_{t_m}\rangle (1-\SC \chi (t-t_{m}))^{2}+ m^{2}\langle \Delta^{2} v_{t_{m}}\rangle \chi^{2}_{x}(t-t_m) + \\
  & + 2m \textrm{Cov}(x_{t_m},v_{t_m})\chi_{x}(t-t_m)(1-\SC \chi (t-t_{m})) \,.
\end{split}
\end{equation}
Note that, because the GLE is linear, if the initial probability distribution function (PDF) $P(x_{t_m},v_{t_m},t_m)$ is a (bivariate) Gaussian, so will be the $P_{t_m}(x_{t},v_{t},t)$ at time $t>t_m$. This also happens if arbitrary initial conditions are taken in the infinite past, i.e. if $t_m \to - \infty$. In fact, if a sufficiently large time has passed between the initial preparation of the system and the observation time $t$, which can be taken positive without loss of generality, the PDF regains its Gaussian character and can hence be written as
\begin{equation}
    P_{t_m}(x_t,v_t,t) = \frac{1}{\sqrt{(2 \pi)^{2}|\CovM_{t_m,t}|}}\exp\left[ -\frac{1}{2}(\textbf{x}_t - \langle \textbf{x} \rangle_{t_m,t}) \CovM_{t_m,t}^{-1} (\textbf{x}_t - \langle \textbf{x} \rangle_{t_m,t}) \right] \, ,
\end{equation}
with $\textbf{x}_t=(x_t,v_t)$, $\langle \textbf{x} \rangle_{t_m,t}=(\langle x \rangle_{t_m,t},\langle v \rangle_{t_m,t})$ and $\CovM_{t_m,t}$ the covariance matrix  
\begin{equation}\label{covmat}
     \CovM_{t_m,t} = 
\begin{pmatrix}
\langle \Delta^{2} x \rangle_{t_m,t} & \textrm{Cov}_{t_m}(x_t,v_t) \\
\textrm{Cov}_{t_m}(x_t,v_t) & \langle \Delta^{2} v \rangle_{t_m,t} \, ,
\end{pmatrix}
\end{equation}
whose components are the variances of position and velocity along with their covariances. We are hence interested in obtaining an expression for the missing components of the covariance matrix:
\begin{equation}\label{varv1}
    \begin{split}
        \langle \Delta^{2}v \rangle_{t_m,t}=& \langle \left( v(t) -\langle v \rangle_{t_m,t}\right)^2 \rangle_{t_m,t}=\\
        =&\partial_{t'} \partial_{t''}\langle \left( x(t') -\langle x \rangle_{t_m,t'}\right)\left( x(t'') -\langle x \rangle_{t_m,t''}\right) \rangle_{t_m,t',t''}\big|_{t'=t''=t} \, ,
    \end{split}
 \end{equation}
 \begin{equation}\label{cov1}
    \begin{split}
        \textrm{Cov}_{t_m}(x_t,v_t)=& \langle x(t) v(t) \rangle_{t_m,t} - \langle x \rangle_{t_m,t} \langle v \rangle_{t_m,t}=\\
        =&\partial_{t'} \langle \left( x(t) -\langle x \rangle_{t_m,t}\right)\left( x(t') -\langle x \rangle_{t_m,t'}\right) \rangle_{t_m,t,t'}\big|_{t'=t} \, ,
    \end{split}
 \end{equation}
 where we used that $\langle v \rangle_{t_m,t} = \partial_{t} \langle x \rangle_{t_m,t}$ because of the linearity of the GLE. Moreover, of course it holds that $\textrm{Cov}_{t_m}(x_t,v_t)=\textrm{Cov}_{t_m}(v_t,x_t)$. \eqref{varv1} and \eqref{cov1} can be computed similarly to the variance of the position \eqref{varx}:
 \begin{equation} \label{varv}
\begin{split}
  \langle \Delta^{2} v \rangle_{t_m,t} =&  k_{B} T \Big[1/m - m \chi^{2}_{v}(t-t_m) - \SC \chi_{x}^{2}(t-t_{m})\Big] + \SC^2 \langle \Delta^{2}x_{t_m}\rangle \chi^{2}_x( t-t_{m})+\\
  &+ m^{2}\langle \Delta^{2} v_{t_{m}}\rangle \chi^{2}_{v}(t-t_m) - 2 \SC m \textrm{Cov}(x_{t_m},v_{t_m})\chi_{v}(t-t_m) \chi_{x} (t-t_{m}) \, ,
\end{split}
\end{equation}
\begin{equation} \label{cov}
\begin{split}
  \textrm{Cov}_{t_m}(x_t,v_t) =& k_{B} T \Big[  \chi_x (t-t_m) - m \chi_{v}(t-t_m)\chi_{x}(t-t_m) - \SC \chi_{x}(t-t_{m})\chi(t-t_{m})\Big]+\\
  & - \SC \langle \Delta^{2}x_{t_m}\rangle \chi_x (t-t_{m}) (1-\SC \chi(t-t_m))+\\
  &+m^{2}\langle \Delta^{2} v_{t_{m}}\rangle \chi_{x}(t-t_m)\chi_{v}(t-t_m) + \\
  & + m \textrm{Cov}(x_{t_m},v_{t_m})\big( \chi_{v}(t-t_m)(1-\SC \chi(t-t_m))- \SC \chi^{2}_{x} (t-t_{m})\big) \, ,
\end{split}
\end{equation}
where we used the convention for the Heaviside step function for which $\theta(0)=1/2$ as well as $\chi_{v}(0)=1/m$. Hence, equations \eqref{varx}, \eqref{varv} and \eqref{cov} are the explicit expressions of the components of the covariance matrix.

\section{Thermodynamic quantities}\label{sec:thermo}

This section is devoted to the analysis of relevant thermodynamic quantities such as work, entropy production and entropy production rate. 

\subsection{Work}

We consider the definition according to stochastic energetics~\cite{sek10,sei12} of  work done on a particle by a time dependent external potential, harmonic in our case, for a particular stochastic trajectory $\omega_{t}$ taking place during the time interval $[0,t]$,
\begin{equation}\label{work1}
\begin{split}
    W (\omega_{t},t) =& - \int_{0}^{t}\mathrm{d}\lambda_{t'} U'(x_{t'}-\lambda(t'))  =\\
    =& -  \SC  \int_{0}^{t} \mathrm{d}\lambda_{t'} (x(t')-\lambda(t'))=\frac{\SC \lambda(t)^2 }{2}-  \SC  \int_{0}^{t}\mathrm{d}\lambda_{t'} x(t') \,
\end{split}    
\end{equation}
where we restricted ourselves to the case where $\lambda(0)=0$. We can calculate the work as a function of the external protocol and the susceptibilities \eqref{chix} and \eqref{chi} by just plugging the explicit solution for the position of the particle \eqref{solution} into \eqref{work1}, which reads
\begin{equation}\label{work2}
\begin{split}
     W (\omega_{t},t) =& \frac{\SC \lambda(t)^2 }{2} - \SC \bigg[  x_{t_m} \left(\lambda(t)-\SC \int_{0}^{t}\mathrm{d}\lambda_{t'}\chi (t'-t_{m})\right) + \\
     & + m \, v_{t_m} \int_{0}^{t}\mathrm{d}\lambda_{t'}\chi_{x} (t'-t_{m}) + \int_{0}^{t}\mathrm{d}\lambda_{t'} \int_{t_m}^{t'}\mathrm{d}t''\chi_{x} (t'-t'') \left[\SC \lambda(t'')+\eta(t'') \right]\bigg]\,.
    \end{split}
\end{equation}
Its average can be obtained, again by noting that $\langle \eta(t) \rangle =0$, as
\begin{equation}\label{avgwork}
\begin{split}
     \langle W \rangle_{t_m,t} = \frac{\SC \lambda(t)^2 }{2} - \SC \bigg[& \langle x_{t_m} \rangle \left(\lambda(t)-\SC \int_{0}^{t}\mathrm{d}\lambda_{t'}\chi (t'-t_{m})\right) +\\&+ m\langle v_{t_m} \rangle \int_{0}^{t}\mathrm{d}\lambda_{t'}\chi_{x} (t'-t_{m}) +\\
     & + \SC \int_{0}^{t}\mathrm{d}\lambda_{t'} \int_{t_m}^{t'}\mathrm{d}t''\chi_{x} (t'-t'') \lambda(t'')\bigg]\,.
    \end{split}
\end{equation}
It is well known that, for such linear systems, the PDF $P(W_t)$ of the work  is Gaussian. In fact, differently from other quantities such as the position, the probability distribution of the work at $t=0$ is always a Dirac delta centred in 0, i.e. $P(W_{t},t=0) = \delta(W_t)$, as it can be easily seen from \eqref{work2}. Since such distribution is the limit of a Gaussian for a random variable with vanishing variance, and given the linearity of the GLE, the PDF of the work stays Gaussian at all times. Hence, in addition to the average $\langle W \rangle_{t_m,t}$, again we need its variance to completely characterise the PDF. It can be calculated similarly to the variance of the position \eqref{Var1}, starting from the definition of work \eqref{work1},
\begin{equation} \label{VarWork1}
\begin{split}
  \langle \Delta^{2} W \rangle_{t_m,t} =& \langle \left(W (x_{t},t) - \langle W \rangle_{t_m,t} \right)^{2} \rangle_{t_m,t}= \SC^2 \bigg \langle \left( \int_{0}^{t} \mathrm{d}\lambda_{t'} ( x(t')-\langle x \rangle_{t_m,t'})\right)^{2} \bigg \rangle_{t_m,t} =\\
  =& \SC^2 \int_{0}^{t}\mathrm{d}\lambda_{t'}\int_{0}^{t}\mathrm{d}\lambda_{t''}\mathcal{C}(t',t'') + \SC^2 \langle \Delta^{2}x_{t_m}\rangle \left(\lambda(t)-\SC \int_{0}^{t}\mathrm{d}\lambda_{t'} \chi (t'-t_{m})\right)^{2}+\\
  &+  m^{2} \SC^2 \langle \Delta^{2} v_{t_{m}}\rangle \left(\int_{0}^{t}\mathrm{d}\lambda_{t'}\chi_x (t'-t_{m}) \right)^{2}  + \\
  & + 2m \SC^2 \textrm{Cov}(x_{t_m},v_{t_m})\left(\lambda_{t}-\SC \int_{0}^{t} \mathrm{d}\lambda_{t'} \chi (t'-t_{m})\right)\int_{0}^{t}\mathrm{d}\lambda_{t'}\chi_x (t'-t_{m}) \, ,
  \end{split}
\end{equation}
where $\mathcal{C}(t',t'')$ was defined in \eqref{Var3}. By computing the first term in the second line we get
\begin{equation} \label{VarWork}
   \begin{split}
       \langle \Delta^{2} W \rangle_{t_m,t} = & k_{B}T \SC^2 \bigg[2 \lambda (t) \int_{0}^{t}\mathrm{d}\lambda_{t'} \chi(t'-t_m) -2 \int_{0}^{t}\mathrm{d}\lambda_{t'}\int_{0}^{t'}\mathrm{d}\lambda_{t''}\chi(t'-t'') + \\
       & -\SC \left( \int_{0}^{t}\mathrm{d}\lambda_{t'}\chi(t'-t_m) \right)^{2} - m \left( \int_{0}^{t}\mathrm{d}\lambda_{t'}\chi_{x}(t'-t_m) \right)^{2} \bigg]+\\
       & + \SC^2 \langle \Delta^{2}x_{t_m}\rangle \left(\lambda(t)-\SC \int_{0}^{t}\mathrm{d}\lambda_{t'} \chi (t'-t_{m})\right)^{2}+\\
        &+ m^{2} \SC^2 \langle \Delta^{2} v_{t_{m}}\rangle \left(\int_{0}^{t}\mathrm{d}\lambda_{t'}\chi_x (t'-t_{m}) \right)^{2}  + \\
        & + 2m \SC^2 \textrm{Cov}(x_{t_m},v_{t_m})\left(\lambda(t)-\SC \int_{0}^{t} \mathrm{d}\lambda_{t'} \chi (t'-t_{m})\right)\int_{0}^{t}\mathrm{d}\lambda_{t'}\chi_x (t'-t_{m}) \, ,
   \end{split} 
\end{equation}
that is the expression for the variance of the work for an arbitrary initial distribution of position and velocities. Although it might look rather complicated, in the next section we will see that the above equation simplifies significantly for some usual initial distributions.

\subsection{Entropy production and entropy production rate}

Entropy production does not need any  introduction, it is a crucial quantity in stochastic thermodynamics that encodes the information about the irreversibility of a given process. In particular, for a colloidal particle in contact with a heat bath, the entropy production for a stochastic trajectory $\omega_{t}$ during a time interval $[0,t]$ can be split into two parts 
\begin{equation}
\Sigma_{\tot}(\omega_{t},t) = \Sigma_{\med}(\omega_{t},t)+\Sigma_{\sys}(\omega_{t},t) 
\end{equation}
with
\begin{equation}
\begin{split}
  \Sigma_{\med}(\omega_{t},t) &=\beta Q (\omega_{t},t) \, , \\ 
  \Sigma_{\sys}(\omega_{t},t) &= -\ln P_{t_m}(x_t,v_t,t) + \ln P_{t_m}(x_{0},v_{0},0) \, ,
\end{split}
\end{equation}
where $Q(\omega_t,t)$ is the heat injected into the heat reservoir, $\beta$ is the inverse temperature (hence $\Sigma_{\med}(x,t)$ is the entropy change in the reservoir) and $\Sigma_{\sys}(x,t)$ is the difference between the Shannon entropy of the final and initial states of the system. In particular, for Gaussian PDFs, it holds that
\begin{equation}\label{EntrSys}
  \Sigma_{\sys}(\omega_{t},t)= \frac{1}{2} \ln \left[ \frac{|\CovM_{t_m,t}|}{|\CovM_{t_m,0}|} \right] \, ,
\end{equation}
where $|\CovM_{t_m,t}|$ is the determinant of the covariance matrix \eqref{covmat} at time $t$.

As for the heat absorbed from the bath, it can be defined through the Stratonovich integral
\begin{equation}\label{heat1}
  Q(\omega_{t},t) = \int_{0}^{t}\mathrm{d}t' F_{bath}(\omega_{t},t') \circ \dot{x}(t') \, ,
\end{equation}
where $F_{bath}(\omega_{t},t)$ is the force exerted from the particle on the bath, i.e., using the GLE \eqref{GLE1},
\begin{align}
    F_{bath}(\omega_{t},t)
    &=\int_{t_{m}}^{t}\mathrm{d}t'\Gamma(t-t') \dot x(t') -\eta (t)\nonumber\\
    &= \SC \lambda(t) -m\ddot x (t) - \SC x (t) \, .
\end{align}
Equation \eqref{heat1} thus becomes 
\begin{align}
  Q(\omega_{t},t) &= \int_{0}^{t}\mathrm{d}t'\left[\SC \lambda(t') -m\ddot x (t') - \SC x (t')\right] \circ \dot{x}(t') = \nonumber\\
  & = \SC \int_{0}^{t}\mathrm{d}t' \lambda(t') \dot{x}(t') - \frac{m}{2}[\dot x^2(t)-\dot x^2(0)] - \frac{\SC}{2}[x^2(t)- x^2(0)] =\nonumber\\
  &= W(\omega_{t},t) -\Delta U (\omega_{t},t) \, ,
\end{align}
where
\begin{equation}
    W(\omega_{t},t) = -  \SC  \int_{0}^{t} \mathrm{d}\lambda_{t'} (x(t')-\lambda(t'))\, ,
\end{equation}
\begin{equation}
    \Delta U (\omega_{t},t) = \frac{m}{2}[ v^2(t)-v^2(0)] + \frac{\SC}{2}[(x(t)-\lambda(t))^2- x^2(0)] \, ,
\end{equation}
recovering the first law of thermodynamics at a stochastic level~\cite{sek10,sei12}.

Taking the average of \eqref{heat1}, as $\langle r^2 \rangle = \langle \Delta^{2}r \rangle + \langle r \rangle^{2} $ for any stochastic variable $r$ and since for underdamped dynamics $\dot x (t) = v(t)$, we get
\begin{equation} \label{EntrMed}
  \begin{split}
   \langle \Sigma_{\med} \rangle_{t_m,t} = \beta \langle Q \rangle_{t_m,t} =&\beta \SC \int_{0}^{t}\mathrm{d}t' \lambda(t') \langle v \rangle_{t_m,t'}+\\
  &- \frac{\beta m}{2}(\langle v\rangle_{t_m,t}^2-\langle v \rangle^{2}_{t_m,0}+\langle \Delta^{2} v \rangle_{t_m,t}-\langle \Delta^{2} v \rangle_{t_m,0}) + \\
  &- \frac{\beta\SC}{2}(\langle x\rangle_{t_m,t}^2-\langle x\rangle^{2}_{t_m,0}+\langle \Delta^{2}x \rangle_{t_m,t}-\langle \Delta^{2} x \rangle_{t_m,0} ) \, .
  \end{split}
\end{equation}
At this stage one can not further simplify this expression for the entropy production. On the other hand, we can obtain a much more compact form for the entropy production rate, defined as
\begin{equation}
    \langle \sigma_{tot} \rangle_{t_m,t} = \partial_{t} \langle \Sigma_{tot} \rangle_{t_m,t}\, .
\end{equation}
For the system entropy production rate we immediately see from \eqref{EntrSys} that 
\begin{equation}\label{EntrSysRate}
    \langle \sigma_{sys} \rangle_{t_m,t} = \frac{\partial_{t}|\CovM_{t_m,t}|}{2|\CovM_{t_m,t}|} \, .
\end{equation}
From equation \eqref{EntrMed} instead we get that 
\begin{equation}\label{EntrMedRate}
  \begin{split} 
  \langle \sigma_{\med} \rangle_{t_m,t} = & \beta \partial_{t}\langle x\rangle_{t_m,t} \left[\SC \lambda(t)- m\partial^{2}_{t}\langle x\rangle_{t_m,t} - \SC \langle x\rangle_{t_m,t} \right] +\\
  &-\frac{\beta \SC}{2} \partial_{t} \langle \Delta x \rangle_{t_m,t}-\frac{\beta m}{2} \partial_{t} \langle \Delta v \rangle_{t_m,t} \, ,\\
  \end{split}
\end{equation}
where again we used that  $\langle v\rangle_{t_{m},t} =\langle \dot{x}\rangle_{t_{m},t}=\partial_{t}\langle x\rangle_{t_{m},t}$. Consider now the term between square brackets on the right hand side of equation \eqref{EntrMedRate} and name it
\begin{equation}
  \mathcal{V}(t,t_m)
  = \SC \lambda(t)- m\partial^{2}_{t}\langle x\rangle_{t_m,t} - \SC \langle x\rangle_{t_m,t} \, .
\end{equation}
Taking its modified Laplace transform we obtain 
\begin{equation}\label{EntrRate1}
\begin{split}
  \mathcal{L}_{k}^{t_m}\left[ \mathcal{V}(t,t_m)\right] =& \SC \hat{\lambda}^{t_m}(k)- \SC \mathcal{L}_{k}^{t_m}\left[\langle x\rangle_{t_m,t}\right] - m k^{2} \mathcal{L}_{k}^{t_m}\left[\langle x\rangle_{t_m,t}\right] \\
  &+ m k \langle x_{t_m}\rangle \e^{-k t_{m}}+ m\langle v_{t_m} \rangle \e^{-k t_{m}} \, ,
\end{split}
\end{equation}
where we used the formula for the modified Laplace transform of a second derivative \eqref{Lap1}. Moreover, looking back to the expression for the average of the position \eqref{avgx} we note that it can be effectively written as
\begin{equation}\label{avgxeff}
   \langle x \rangle_{t_m,t} = \mathcal{I}(t,t_m)+\SC \int^{t}_{t_m} \mathrm{d}t' \chi_{x} (t-t') \lambda(t') \, ,
\end{equation}
where $\mathcal{I}(t,t_m) = \langle x_{t_m} \rangle(1-\SC \chi (t-t_{m}))+ m\langle v_{t_{m}}\rangle \chi_{x}(t-t_m)$  contains the information relative to initial conditions, in particular  $\mathcal{I}(t_m,t_m)=\langle x_{t_m} \rangle$.

Going back to equation \eqref{EntrRate1}, recalling the definition of the position susceptibility via its Laplace transform
($\hat{\chi}_{x}(k) = [m k^{2}+k\hat{\Gamma}(k)+\SC]^{-1}$) and using that for the first term on the right hand side of equation \eqref{avgxeff} we have that 
\begin{equation}
  \hat{\mathcal{I}}^{t_m}(k) = \langle x_{t_m} \rangle \frac{\e^{-k t_{m}}}{k} (1-\SC \hat{\chi}_{x}(k)) + m \langle v_{t_m} \rangle \e^{-k t_{m}} \hat{\chi}_{x}(k) \, ,
\end{equation}
along with the generalised convolution theorem for the second one, we get
\begin{equation}\label{EntrRate2}
\begin{split}
  \mathcal{L}_{k}^{t_m}\left[ \mathcal{V}(t,t_m)\right] =& \SC \hat{\lambda}^{t_m}(k)- \SC \mathcal{L}_{k}^{t_m}\left[\langle x\rangle_{t_m,t}\right] - m k^{2} \mathcal{L}_{k}^{t_m}\left[\langle x\rangle_{t_m,t}\right]+\\
  & + m k \langle x_{t_m}\rangle\e^{-k t_{m}} + m\langle v_{t_m} \rangle \e^{-k t_{m}}=\\
  =& \SC \left[ 1 - \SC \hat{\chi}_{x}(k) -m k^{2}\hat{\chi}_{x}(k)\right]\hat{
  \lambda}^{t_m}(k)-(m k^{2}+\SC)\hat{\mathcal{I}}^{t_m}(k)+\\
  & + m k \langle x_{t_m}\rangle\e^{-k t_{m}} + m\langle v_{t_m} \rangle \e^{-k t_{m}}=\\
  =&  \hat{\Gamma}(k)\left[ \SC k \hat{\chi}_{x}(k) \hat{\lambda}^{t_m}(k)+k\hat{\mathcal{I}}^{t_m}(k)- \mathcal{I}(t_m,t_m) \e^{-k t_{m}}\right]=\\
  =& \hat{\Gamma}(k) \mathcal{L}_{k}^{t_m}\left[\partial_{t}\langle x\rangle_{t_m,t}\right] \, .
\end{split}
\end{equation}
Its inverse can be calculated using again the convolution theorem
\begin{equation}\label{RetVel}
    \mathcal{V}(t,t_m) = \int_{t_m}^{t}\mathrm{d}t'\Gamma(t-t') \langle v\rangle_{t_m,t'}=  \int^{t-t_m}_{0}\mathrm{d}t' \langle v\rangle_{t_m,t-t'} \Gamma(t') = \fric(t-t_m) \langle v_{\text{ret}}\rangle_{t_m,t}\, ,
\end{equation}
where 
\begin{align}
    \fric(t) = \int_{0}^{t}\mathrm{d}t'\Gamma(t') 
\end{align}
is the time dependent effective friction coefficient and $\fric =  \displaystyle \lim_{t \to \infty} \fric(t)$ is its asymptotic limit for long times. Moreover, we define the \textit{retarded velocity} as 
\begin{equation} \label{ret_vel}
    \langle v_{\text{ret}}\rangle_{t_m,t} =  \frac{1}{\fric(t-t_m)} \int^{t-t_m}_{0}\mathrm{d}t' \langle v\rangle_{t_m,t-t'} \Gamma(t')
\end{equation}
which can be interpreted as a quantity converging to the real velocity for $t \to \infty$, i.e.
\begin{equation} \label{limitvel}
\begin{split}
    \displaystyle \lim_{t \to \infty} \langle v_{\text{ret}} \rangle_{t_m,t} =& \displaystyle \lim_{t \to \infty} \frac{1}{\fric(t-t_m)} \int^{t-t_m}_{0}\mathrm{d}t' \langle v\rangle_{t_m,t-t'} \Gamma(t') \approx\\
    \approx & \displaystyle \lim_{t \to \infty} \frac{\langle v\rangle_{t_m,t}}{\fric(t-t_m)} \int^{t-t_m}_{0}\mathrm{d}t'  \Gamma(t')= \lim_{t \to \infty} \langle v\rangle_{t_m,t} \, .
\end{split}
\end{equation}
The same decoupling between the kernel and the average velocity can be obtained for $t_m\to -\infty$ if one is able to show that $\langle v\rangle_{t_m,t}=\langle v\rangle_{t-t_m}$. It will be for example the case of a trapped particle dragged at a constant velocity, i.e. $\lambda(t)=vt$. In fact, under these hypothesis and with a calculation analogous to that of equation \eqref{limitvel}, we see that
\begin{equation} 
\begin{split}
    \displaystyle \lim_{t_m \to -\infty} \langle v_{\text{ret}} \rangle_{t-t_m} =& \displaystyle \lim_{t_m \to -\infty} \frac{1}{\fric(t-t_m)} \int^{t-t_m}_{0}\mathrm{d}t' \langle v\rangle_{t-t_m-t'} \Gamma(t') = \lim_{t_m \to -\infty} \langle v\rangle_{t-t_m}\,.
\end{split}
\end{equation}
Moreover, note that for Markovian dynamics defined by a memory kernel $\Gamma^{\text{Markov}}(t) = 2\gamma_{0} \delta (t)$ it holds that $\fric = \fric(t)= \gamma_{0}$ and $\langle v_{\text{ret}}\rangle_{t_m,t} = \langle v \rangle_{t_m,t}$ for every $t$. 

Finally, putting together equation \eqref{EntrMedRate} and \eqref{RetVel}, we get 
\begin{equation}
  \langle \sigma_{\med} \rangle_{t_m,t} = \beta \fric(t-t_m) \langle v\rangle_{t_m,t} \langle v_{\text{ret}}\rangle_{t_m,t} -\frac{\beta \SC}{2} \partial_{t} \langle \Delta^{2} x \rangle_{t_m,t} -\frac{\beta m }{2} \partial_{t} \langle \Delta^{2} v \rangle_{t_m,t} 
\end{equation}
while for the total entropy production rate (assuming that $P_{t_m}(x_t,v_t,t)$ is Gaussian) we have that
\begin{equation}\label{avgEntrTot}
  \langle \sigma_{\tot} \rangle_{t_m,t} = \beta \fric(t-t_m) \langle v\rangle_{t_m,t} \langle v_{\text{ret}}\rangle_{t_m,t} -\frac{\beta \SC}{2} \partial_{t} \langle \Delta^{2} x \rangle_{t_m,t}  -\frac{\beta m }{2} \partial_{t} \langle \Delta^{2} v \rangle_{t_m,t} + \frac{\partial_{t}|\CovM_{t_m,t}|}{2|\CovM_{t_m,t}|}\, .
\end{equation}

\section{Overdamped dynamics}\label{sec:over}

Until now we restricted our discussion to underdamped dynamics, namely considering a finite mass for the particle and hence including inertial effects in the GLE \eqref{GLE2}. Instead, the overdamped case can be considered by taking $m=0$, corresponding to the following GLE
\begin{equation}\label{GLE_over}
     \int_{t_m}^{t}\mathrm{d}t'\Gamma(t-t') \dot x (t') =  - \SC [x(t)-\lambda(t)] + \eta (t)\, .
\end{equation}
Its solution can be obtained with the same procedure used for the underdamped case with the main difference consisting in a different definition of the position susceptibility 
\begin{equation} \label{chix_over}
\hat{\chi}^{\text{over}}_{x}(k) = [k\hat{\Gamma}(k)+\SC]^{-1}
\end{equation}
and, as a consequence, of the other susceptibilities
\begin{align} \label{chi_over}
  \chi^{\text{over}}(t) &\equiv \int_{0}^{t} \mathrm{d}t' \chi^{\text{over}}_{x}(t')\, , \\
  \chi^{\text{over}}_{v}(t) &\equiv \partial_{t} \chi^{\text{over}}_{x}(t) \, .
\end{align}
It is important to underline that one can not explicitly calculate the underdamped susceptibilities and take the massless limit $m \to 0$ afterwards because this would lead to inconsistencies, as it can be seen in \cite{LoosKlapp_heat_flow}. 
However, the direct solution of the overdamped dynamics \eqref{GLE_over} can be found (dropping the "over" superscript):
\begin{equation}\label{solution_over}
   x(t) = x_{t_m} \left( 1 - \SC \chi (t-t_m) \right) + \int_{t_m}^{t}\mathrm{d}t'\chi_{x} (t-t') \left[\SC \lambda(t')+\eta(t') \right] 
\end{equation}
with its average equal to
\begin{equation}\label{xavg_over}
   \langle x \rangle_{t_m,t} = \langle x_{t_m} \rangle \left( 1 - \SC \chi (t-t_m) \right) + \SC \int_{t_m}^{t}\mathrm{d}t'\chi_{x} (t-t') \lambda(t') 
\end{equation}
and with variance
\begin{equation} \label{varx_over}
\begin{split}
  \langle \Delta^{2} x \rangle_{t_m,t} = k_{B} T \Big[ 2 \chi(t-t_m) - \SC \chi^{2}(t-t_{m}) \Big] + \langle \Delta^{2}x_{t_m}\rangle (1-\SC \chi (t-t_{m}))^{2} \, .
\end{split}
\end{equation}
The velocity is computed by taking the derivative of \eqref{solution_over},
\begin{equation}\label{solution_velocity_over}
   v(t) = - \SC x_{t_m} \chi_{x}(t-t_m)  + \int_{t_m}^{t}\mathrm{d}t'\chi_{v} (t-t') \left[\SC \lambda(t')+\eta(t') \right] + \chi_{x}(0) \left[\SC \lambda(t)+\eta(t) \right ] \, .
\end{equation}
Since in the overdamped case $\chi_{x}(0) \neq 0$ (see Appendix \ref{sec:limits}), the velocity is proportional to the noise $\eta(t)$, corresponding to the well known singularity of Brownian motion. This feature disappears once the average is taken,
\begin{equation}\label{vavg_over}
   \langle v \rangle_{t_m,t} = - \SC \langle x_{t_m} \rangle \chi_{x}(t-t_m)  + \SC \int_{t_m}^{t}\mathrm{d}t'\chi_{v} (t-t') \lambda(t') + \SC \chi_{x}(0) \lambda(t') \, .
\end{equation}
On the other hand, the variance of the velocity is not well defined as the $\chi_{x}(0)\eta(t)$ term again yields some mathematical problems. Indeed, trying to calculate this variance, one finds a term of the form $\chi^{2}_{x}(0)\langle \eta(t) \eta(t) \rangle= k_{B} T \chi^{2}_{x}(0) \Gamma(0) $, which is a singular quantity (consider Markov dynamics for example), see again Appendix \ref{sec:limits} for more details.

As for the work and its variance, again making the same reasoning as above, we get
\begin{equation}\label{avgwork_over}
\begin{split}
     \langle W \rangle_{t_m,t} = \frac{\SC \lambda(t)^2 }{2} - \SC \bigg[& \langle x_{t_m} \rangle \left(\lambda(t)-\SC \int_{0}^{t}\mathrm{d}\lambda_{t'}\chi (t'-t_{m})\right) +\\
     & + \SC \int_{0}^{t}\mathrm{d}\lambda_{t'} \int_{t_m}^{t'}\mathrm{d}t''\chi_{x} (t'-t'') \lambda(t'')\bigg] \, ,
    \end{split}
\end{equation}
\begin{equation} \label{VarWork_over}
   \begin{split}
       \langle \Delta^{2} W \rangle_{t_m,t} = & k_{B}T \SC^2 \bigg[2 \lambda (t) \int_{0}^{t}\mathrm{d}\lambda_{t'} \chi(t'-t_m) -2 \int_{0}^{t}\mathrm{d}\lambda_{t'}\int_{0}^{t'}\mathrm{d}\lambda_{t''}\chi(t'-t'') + \\
       & -\SC \left( \int_{0}^{t}\mathrm{d}\lambda_{t'}\chi(t'-t_m) \right)^{2} \bigg]+ \SC^2 \langle \Delta^{2}x_{t_m}\rangle \left(\lambda(t)-\SC \int_{0}^{t}\mathrm{d}\lambda_{t'} \chi (t'-t_{m})\right)^{2} \, .
   \end{split} 
\end{equation}
Finally, for a Gaussian PDF, obtained for example starting from equilibrium initial conditions or by sending $t_m \to - \infty$ and $t \ge 0$, we get the following expressions for the total entropy production rate
\begin{equation}\label{avgEntrTot_over}
  \langle \sigma_{\tot} \rangle_{t_m,t} = \beta \fric(t-t_m) \langle v\rangle_{t_m,t} \langle v_{\text{ret}}\rangle_{t_m,t} -\frac{\beta \SC}{2} \partial_{t} \langle \Delta^{2} x \rangle_{t_m,t}  + \frac{\partial_{t}\langle \Delta^{2} x \rangle_{t_m,t}}{2\langle \Delta^{2} x \rangle_{t_m,t}} \, .
\end{equation}

\section{Applications}\label{sec:appl}

In this paragraph we apply the general formulas derived in the previous sections to specific initial conditions. In particular, we will discuss two cases:
\begin{itemize}
\item Dynamics starting from equilibrium conditions, generated by a trap left still for a long time with its minimum at $x=0$, implying that $\langle x_{0} \rangle_{t}^{\text{eq}}=0$ and $\langle v_{0} \rangle_{t}^{\text{eq}}=0$. The protocol starts at $t=0$ and no memory with the past is established, meaning that $t_m=0$.
\item Dynamics starting in the infinite past, corresponding to $t_m \to - \infty$, where memory of initial conditions is lost. Moreover, we will show that for the particular case of a linear dragging protocol $\lambda(t)=vt$, the system reaches a nonequilibrium steady state. This happens because the system can be mapped, through a Galileian transformation, to a reference frame where an equilibrium distribution is achieved in the limit $t_m \to - \infty$.
\end{itemize}

Of course, for a given protocol, in both cases the dynamics of the system becomes the same in the limit of large observation times $t \to \infty$. 

Moreover, we stress that all the formulae presented in this section are both valid for underdamped and overdamped dynamics, with the only difference that the susceptibilities must be calculated at the beginning by choosing respectively a finite or a null mass for the particle.

\subsection{Dynamics starting from equilibrium}

For a colloidal particle trapped in a parabolic potential with stiffness $\SC$,
the equilibrium PDF at time $t_m=0$ has a Gaussian shape, 
\begin{equation}
     P^{\text{eq}}(x_0,v_0) = \frac{1}{\sqrt{(2 \pi)^{2}|\CovM^{\text{eq}}_{0}|}}\exp\left[ -\frac{1}{2}(\textbf{x}_0 - \langle \textbf{x}_{0} \rangle^{\text{eq}}) \left( \CovM_{0}^{\text{eq}}\right)^{-1} (\textbf{x}_0 - \langle \textbf{x}_{0} \rangle^{\text{eq}}) \right] \, ,
\end{equation}
with parameters given by
\begin{equation}\label{xveqin}
    \langle \textbf{x}_{0} \rangle^{\text{eq}}=
\begin{pmatrix}
\langle x_0 \rangle^{\text{eq}} \\
\langle v_0 \rangle^{\text{eq}}  
\end{pmatrix}=
\begin{pmatrix}
0 \\
0 
\end{pmatrix} \, ,
\end{equation}
\begin{align}\label{covmateqin}
     \CovM^{eq}_{0} = 
\begin{pmatrix}
\langle \Delta^{2} x_0 \rangle^{\text{eq}} & \textrm{Cov}^{\text{eq}}(x_0,v_0) \\
\textrm{Cov}^{\text{eq}}(x_0,v_0) & \langle \Delta^{2} v_{0} \rangle^{\text{eq}} 
\end{pmatrix}=
\begin{pmatrix}
\frac{k_{B}T}{\SC} & 0 \\
0 & \frac{k_{B}T}{m} 
\end{pmatrix} \, .
\end{align}
Using equations \eqref{xveqin} and \eqref{covmateqin}, we can evaluate the evolution of all the quantities discussed in the previous section, starting from the probability distribution defined above and for an arbitrary $\lambda(t)$. Starting from the average of the position  \eqref{avgx} and velocity \eqref{avgv} we find that 
\begin{equation}\label{xveq}
    \langle \textbf{x} \rangle^{\text{eq}}_{t}=
\begin{pmatrix}
\langle x \rangle^{\text{eq}}_t \\[7pt]
\langle v \rangle^{\text{eq}}_t  
\end{pmatrix}= \SC
\begin{pmatrix}
\int^{t}_{0}\mathrm{d}t' \chi_{x} (t-t') \lambda(t')\\[7pt]
\int^{t}_{0}\mathrm{d}t' \chi_{v} (t-t') \lambda(t')
\end{pmatrix} 
\end{equation}
while for the covariance matrix, using equations \eqref{varx}, \eqref{varv} and \eqref{cov} we get that
\begin{align}\label{covmateq}
     \CovM^{eq}_{t} = 
\begin{pmatrix}
\langle \Delta^{2} x \rangle^{\text{eq}}_t & \textrm{Cov}^{\text{eq}}(x_t,v_t) \\
\textrm{Cov}^{\text{eq}}(x_t,v_t) & \langle \Delta^{2} v \rangle^{\text{eq}}_t 
\end{pmatrix}=
\begin{pmatrix}
\frac{k_{B}T}{\SC} & 0 \\
0 & \frac{k_{B}T}{m} 
\end{pmatrix}\, ,
\end{align}
i.e.~if we start from equilibrium and the trap stiffness $\SC$ does not change, then the covariance matrix remains constant in time for every choice of $\lambda(t)$.

Going forward to the estimate of thermodynamic work, from \eqref{avgwork} and \eqref{VarWork} and again using that $\lambda(0) =0$ along with $\chi(t)=\int_{0}^{t}\mathrm{d}t'\chi_{x}(t)$, we get that
\begin{equation}\label{avgworkeq}
    \langle W \rangle_{t}^{eq} = \SC \left(\frac{ \lambda(t)^2}{2}-  \SC \int_{0}^{t}\mathrm{d\lambda_{t'}} \int_{0}^{t'} \mathrm{d\lambda_{t''}} \chi(t'-t'')\right) \, ,
\end{equation}
\begin{equation}\label{VarWorkeq}
    \langle \Delta^{2} W \rangle^{eq}_{t} = 2 k_B T \SC \left( \frac{ \lambda(t)^2}{2}-  \SC \int_{0}^{t}\mathrm{d\lambda_{t'}} \int_{0}^{t'} \mathrm{d\lambda_{t''}} \chi(t'-t'')\right) \, ,
\end{equation}
i.e.
\begin{equation}\label{genVZC}
    \langle \Delta^{2} W \rangle^{eq}_{t} = 2 k_B T \langle W \rangle^{eq}_{t} \, .
\end{equation}
Since the PDF of the work $P(W_t)$ is Gaussian, an integral fluctuation theorem for the thermodynamic work $W(x_t,v_t,t)$ holds (see \cite{WorkFT_NonMarkov} for details) and a Jarzynski equality would follow~\cite{Speck_Seifert_FT_non-Markov}.

Finally, since the covariance matrix and its determinant are both constants, a very simple expression can be found for the rate of entropy production 
\begin{equation}
    \langle \sigma_{\med} \rangle^{\text{eq}}_{t} = \frac{ \fric(t) \langle v\rangle^{\text{eq}}_{t} \langle v_{\text{ret}}\rangle^{\text{eq}}_{t}}{k_B T}\, ,
\end{equation}
where again
\begin{equation} \label{ret_vel_eq}
    \langle v_{\text{ret}}\rangle_{t}^{\text{eq}} =  \frac{1}{\fric(t)} \int^{t}_{0}\mathrm{d}t' \langle v\rangle_{t-t'}^{\text{eq}} \Gamma(t') \, .
\end{equation}

\subsection{Initial conditions in the infinite past}

We discuss the evolution of all the quantities presented in the previous sections when the initial conditions are taken in the infinite past, i.e. $t_m \to - \infty$. This can be considered as a "stationary state" in a generalised sense, meaning that memory of initial conditions is lost and, as we will see in few lines, that the covariance matrix has become constant. This can be easily seen again by considering the limits of the susceptibilities discussed in the appendix. For position and velocity, using again equations \eqref{avgx} and \eqref{avgv}, what we get is
\begin{equation}\label{xvssg}
     \langle \textbf{x} \rangle_{-\infty,t}=  \SC
\begin{pmatrix}
\int^{t}_{-\infty}\mathrm{d}t' \chi_{x} (t-t') \lambda(t')\\[7pt]
\int^{t}_{-\infty}\mathrm{d}t' \chi_{v} (t-t') \lambda(t')
\end{pmatrix}\, .
\end{equation}
As for the covariance matrix, we again use the expressions for the variance of position and velocity \eqref{varx} and \eqref{varv} alongside with their covariance \eqref{cov}, finding that
\begin{align}\label{covmatss}
     \displaystyle\lim_{t_m \to -\infty}\CovM_{-\infty,t} = 
\begin{pmatrix}
\langle \Delta^{2} x \rangle_{-\infty,t} & \textrm{Cov}_{-\infty}(x_t,v_t) \\[5pt]
\textrm{Cov}_{-\infty}(x_t,v_t) & \langle \Delta^{2} v \rangle_{-\infty,t}
\end{pmatrix}=
\begin{pmatrix}
\frac{k_{B}T}{\SC} & 0 \\[5pt]
0 & \frac{k_{B}T}{m} 
\end{pmatrix} \, .
\end{align}
As in the previous example starting form equilibrium, also for this sort of steady state we have that the covariance matrix does not depend on time for every driving protocol $\lambda(t)$.

The average work can be readily calculated using that $\chi(\infty)=1/\SC$ along with $\chi_{x}(\infty)=0$, namely
\begin{equation}\label{avgworkssg}
     \langle W \rangle_{-\infty,t} = \SC \left(\frac{ \lambda(t)^2 }{2} - \SC \int_{0}^{t}\mathrm{d}\lambda_{t'} \int_{-\infty}^{t'}\mathrm{d}t''\chi_{x} (t'-t'') \lambda(t'') \right)\, .
\end{equation}
As for its variance instead, we obtain that
\begin{equation} \label{varWeqstaz}
    \langle \Delta^{2} W \rangle_{-\infty,t} = \langle \Delta^{2} W \rangle^{eq}_{t} = 2 k_B T \SC \left( \frac{ \lambda(t)^2}{2}-  \SC \int_{0}^{t}\mathrm{d\lambda_{t'}} \int_{0}^{t'} \mathrm{d\lambda_{t''}} \chi(t'-t'')\right)
\end{equation}
i.e. the variance of the work in the generalised steady state is equal to the one starting from equilibrium conditions \eqref{VarWorkeq} for every driving protocol $\lambda(t)$. 

Finally, for the entropy production rate we use equation \eqref{avgEntrTot} along with the fact that the covariance matrix is constant in order to obtain 
\begin{equation}
  \langle \sigma_{\tot} \rangle_{-\infty,t} = \frac{ \fric \langle v\rangle_{-\infty,t} \langle v_{\text{ret}}\rangle_{-\infty,t} }{k_B T}\, ,
\end{equation}
with
\begin{equation} \label{ret_vel_gen_steady_gen}
    \langle v_{\text{ret}}\rangle_{-\infty,t} =  \frac{1}{\fric} \int^{\infty}_{0}\mathrm{d}t' \langle v\rangle_{-\infty,t-t'}\Gamma(t') \, .
\end{equation}

\subsubsection{Steady state}

A particularly interesting case to consider is a linear dragging protocol of the form $\lambda(t)=vt$, where a nonequilibrium steady state is reached in the limit $t_m \to - \infty$. To understand why this happens, we recall that one usually defines the stationary distribution as the solution of the Fokker-Planck equation when the PDF does not depend explicitly on time. Nevertheless, this definition becomes problematic when the drift term or the diffusion coefficient of the associated Langevin equation depend explicitly on time, as in the cases we are considering in this paper. To tackle this problem, first of all we note that if a sufficiently large time has passed from the beginning of the dynamics, i.e. if $t_m \to - \infty$, the PDF $P_{t_m}(x_t,v_t,t)$ at time $t \ge 0$ will be a bivariate Gaussian with the usual form
\begin{equation}
     \displaystyle\lim_{t-t_m \to +\infty}P_{t_m}(x_t,v_t,t) = \frac{1}{\sqrt{(2 \pi)^{2}|\CovM_{t_m,t}|}}\exp\left[ -\frac{1}{2}(\textbf{x}_t - \langle \textbf{x} \rangle_{t_m,t}) \CovM_{t_m,t}^{-1} (\textbf{x}_t - \langle \textbf{x} \rangle_{t_m,t}) \right]\, ,
\end{equation}
depending on time via the averages of position and velocity and the covariance matrix. From \eqref{covmatss} we see that for initial conditions taken in the infinite past the covariance matrix does not depend on time for every driving protocol $\lambda(t)$, but this does not happen in general for the averages of position of velocity, as it can be seen from equation \eqref{xvssg}.

We outflank this problem by moving the centre of the harmonic trap at constant speed, i.e.~$\lambda(t)=vt$, so that we get the following GLE
\begin{equation} 
 m\ddot x (t) = - \int_{t_{m}}^{t}\Gamma(t-t') \dot x(t') \mathrm{d}t'- \SC \left[ x (t) - v t\right]  + \eta (t) \, .
\end{equation}
Performing the change of variable $y(t)=x(t)-vt$, we see that the system can be mapped through a Galilean transformation to the centre of the trap reference frame. This is always a consistent procedure for a GLE, as shown in \cite{Cairoli_Galileian_invariance}. Moreover, note that this transformation does not change the covariance matrix and that the new PDF $P_{t_m}(y_t,\dot y_t,t)$ will be defined by the same matrix along with $\langle y \rangle_{t_m,t}$ and $\langle \dot y \rangle_{t_m,t}$, which we will be now explicitly calculated. The transformed GLE hence becomes
\begin{equation} 
 m\ddot y (t) = - \int_{t_{m}}^{t}\Gamma(t-t') \dot y(t') \mathrm{d}t' - v\int_{t_{m}}^{t}\Gamma(t-t') \mathrm{d}t' - \SC y (t)  + \eta (t)
\end{equation}
and its solution can be found similarly to that for the original GLE. In particular we find that
\begin{equation}
  \langle y \rangle _{t_m,t} = \langle y_{t_m} \rangle (1 - \SC \chi (t-t_m)) + m \langle \dot{y}_{t_m} \rangle \chi_{x}(t-t_m) - v \int_{0}^{t-t_m}\mathrm{d}t'\chi (t-t_m-t')\Gamma(t') \, .
\end{equation}
Taking the limit $t_m \to - \infty$ and using the limits derived in Appendix~\ref{sec:limits}, we see that
\begin{align} \label{limy}
  \displaystyle\lim_{t_m \to - \infty}\langle y \rangle_{t_m,t} &
  = - v\chi(\infty) \int_{0}^{\infty}\Gamma(t') \mathrm{d}t' =-\frac{\fric v}{\SC} \, , \nonumber\\
  \displaystyle\lim_{t_m \to - \infty}\langle \dot y \rangle_{t_m,t} &
  = 0 \, ,
\end{align}
which are both constant.
We conclude that for a harmonic potential with constant strength and with centre travelling at constant speed ($\lambda(t)=vt$) it is possible, through a Galilean transformation, to map the system to another one for which an equilibrium distribution exists. In fact, the PDF $P_{t_m}(y_t,\dot y_t,t)$ inherits the Gaussian character from the PDF of the original variable $x(t)$.
Thus, the PDF for $y(t)$ becomes time independent because the covariance matrix and the averages of the dynamical variables \eqref{limy} are constant.
In this sense we mean that $P_{t_m}(x_t,v_t,t)$ becomes stationary as $t_m \to - \infty$. 

Introducing now the notation $\langle \cdot \rangle^{ss}$, meaning that we are considering stationary averages in the sense discussed above, we note that
\begin{align}\label{avgxss}
    \langle x \rangle^{\text{ss}}_{t}   =vt+\displaystyle\lim_{t_m \to - \infty}\langle y \rangle_{t_m,t}  =vt-\frac{\fric v}{\SC}\,, &&  \langle v \rangle^{\text{ss}}_{t} = v \, ,
\end{align}
i.e.~they do not depend on the specific form of the memory kernel but only on the limit of its time integral. Moreover, note that the expressions above exhibit the same structure as in the usual Markov case where instead of $\fric$ there appears the conventional Stokes friction coefficient $\gamma_0$. 

Consider now the thermodynamic work, in particular equations \eqref{avgwork} and \eqref{VarWork} for the specific case of $\lambda(t)=vt$. For the average work we find
\begin{equation}
\begin{split}
    \langle W \rangle^{\text{ss}}_{t} =\frac{\SC v^2 t^2 }{2}-  \SC v  \int_{0}^{t}\mathrm{d}t' \langle x \rangle^{\text{ss}}_{t'} = \fric v^2 t \, ,
\end{split}    
\end{equation}
that again has the same form as the well known Markov case. For the variance of the work, instead, we use the limits of susceptibilities discussed in Appendix~\ref{sec:limits}, hence obtaining
\begin{equation}
    \langle \Delta^{2} W \rangle_{t}^{\text{ss}} =  k_{B}T\SC v^2\left(t^2- 2 \SC  \int_{0}^{t}\mathrm{dt'}\int_{0}^{t'}\mathrm{dt''} \chi(t'') \right) \, .
\end{equation}
As for the entropy production rate we immediately see that it has the same form as for Markov dynamics with the usual substitution $\gamma_0 \to \fric$
\begin{equation}
    \langle \sigma_{tot} \rangle_{t}^{\text{ss}} = \frac{ \fric \langle v\rangle^{\text{ss}}_{t} \langle v_{\text{ret}}\rangle^{\text{ss}}_{t} }{k_B T}
 = \fric v^2 \, ,
\end{equation}
because
\begin{equation} \label{ret_vel_gen_steady}
    \langle v_{\text{ret}}\rangle^{\text{ss}}_{t} =  \frac{1}{\fric} \int^{\infty}_{0}\mathrm{d}t' \langle v\rangle_{t-t'}^{\text{ss}} \Gamma(t') = \frac{v}{\fric} \int^{\infty}_{0}\mathrm{d}t' \Gamma(t') = v \, .
\end{equation}
Moreover, the constancy of the entropy production rate is another indicator that the scenario discussed above is indeed a stationary state.

\subsection{Example: exponentially decaying memory kernel}

As a standard example for non-Markovian dynamics, we examine a GLE with exponentially decaying memory kernel, as in Maxwell model for viscoelasticity~\cite{tas16ch6}. In particular, we examine two cases:
underdamped dynamics and overdamped dynamics.
For causality, in both cases it holds that the memory kernel $\Gamma^{\text{exp}}(t<0) = 0$.

\subsubsection{Underdamped dynamics}

\begin{figure}[tb]
\includegraphics[width=0.95\textwidth, angle=0]{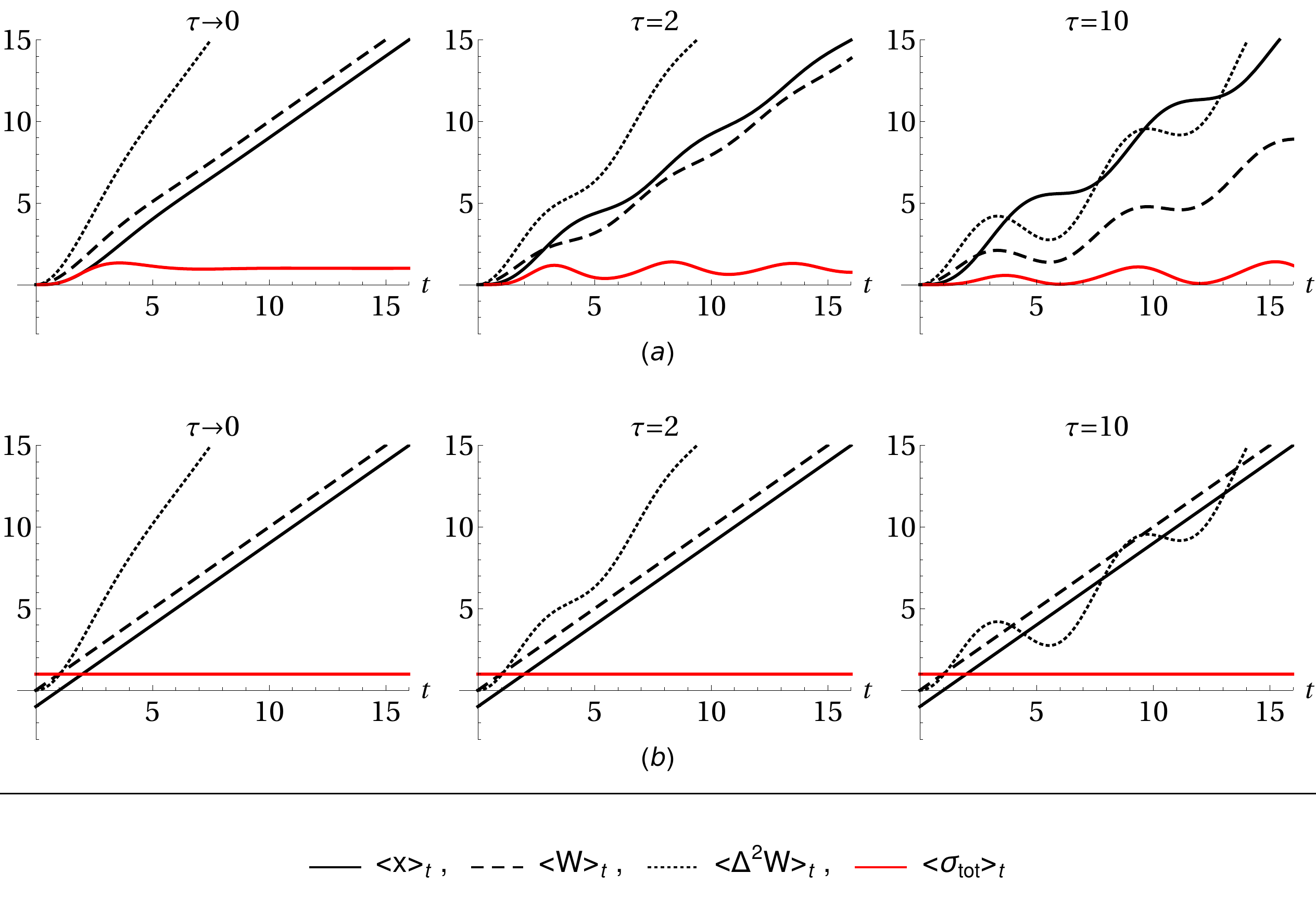}
\caption{Time evolution of some of the quantities discussed in the previous sections starting from equilibrium (a) and from a stationary state (b) for linear dragging protocol $\lambda(t)=vt$. Parameters are set as $m=1$, $\SC=1$, $\gamma=1$ and $v=1$. For (a) we see that as $\tau$ increases oscillations arise for all quantities while for (b) oscillations are visible only for $\langle \Delta W \rangle^{ss}_{t}$ as it is equal to $\langle \Delta W \rangle^{eq}_{t}$. Moreover, note that for the second column (i.e. $\tau=2$), the effects of memory are still very present even at an observation time $t$ equal to several multiples of $\tau$. }
\label{Fig2}
\end{figure}
\begin{figure}[tb]
\includegraphics[width=0.65\textwidth, angle=0]{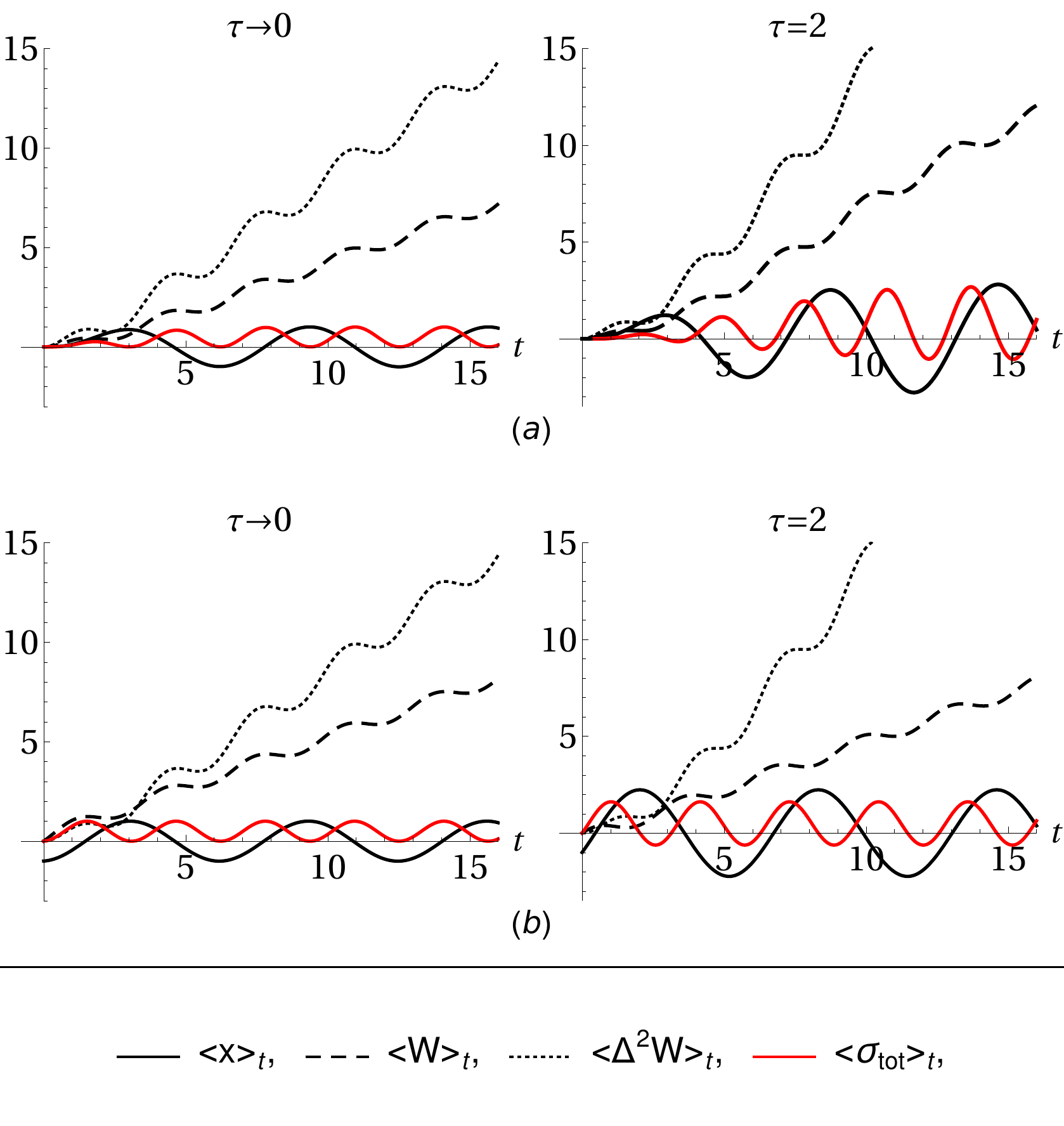}
\caption{Evolution of the same thermodynamic quantities as in the previous figure starting from equilibrium (a) and for $t_m \to - \infty$ (b) for dragging protocol $\lambda(t)= A sin(\omega t)$. We set $m=1$, $\SC=1$, $\gamma=1$, $A=1$ and $\omega=1$. In both scenarios we observe an increasing amplitude of the oscillations that are already present because of the intrinsic oscillatory nature of the driving protocol. This is particularly evident for the average of the position. Another interesting feature that can be observed is that the entropy production rate can become negative as memory effects arise. Note that even in this case, differences between the two columns are still present at an observation time $t$ much larger then $\tau$.}
\label{Fig3}
\end{figure}

\begin{figure}[tb]
\includegraphics[width=0.78\textwidth, angle=0]{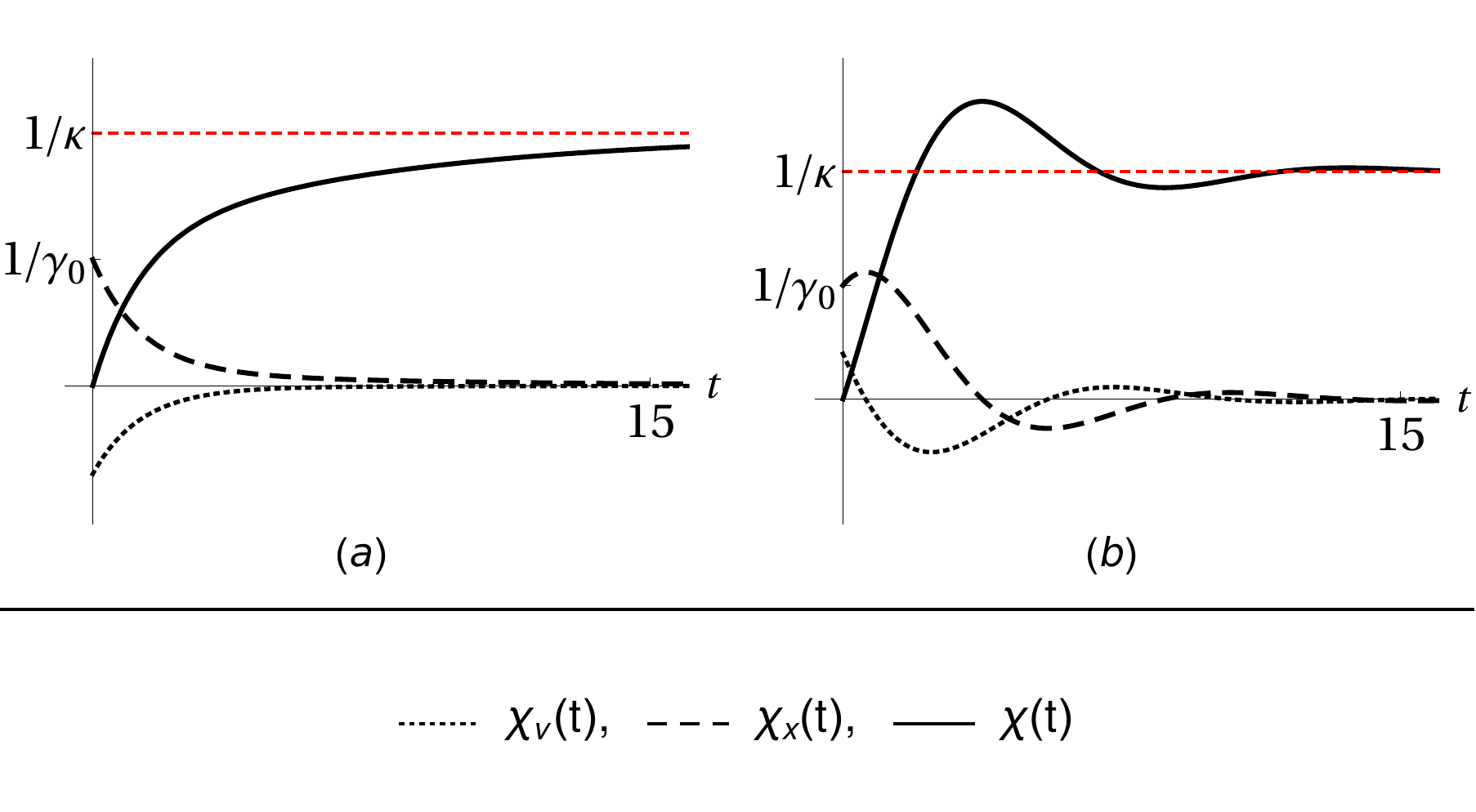}
\caption{Overdamped ($m=0$) susceptibilities for memory kernel of the form given in equation \eqref{KernelExpOver2}. For both figures we set $\kappa = 1$ and $\gamma_0 =1$ while for the exponential part of the kernel we chose (a) $\gamma=1$, $\tau=5$ and (b) $\gamma=-0.9$, $\tau=1$. The limits of the susceptibilities coincide with those calculated in Appendix~\ref{sec:limits}. Note that oscillations only appear in the case where the exponential part of the kernel is negative.}
\label{Fig4}
\end{figure}

We first discuss the underdamped GLE with a purely exponential memory kernel
\begin{equation}
    \Gamma^{\text{exp}}(t) = \frac{\gamma}{\tau} \exp[-t/\tau]\quad\textrm{for } t\ge 0 \, .
\end{equation}
The characteristic time $\tau$ could emerge, for example, from the relaxation of the molecules or polymers in the reservoir.
In the limit $\tau \to 0$, the symmetrized memory kernel tends to twice the Dirac delta 
\begin{equation}\label{KernelExpLimit}
    \displaystyle \lim_{\tau \to 0} \Gamma^{\text{exp}}(|t|) = 2 \gamma \delta (t)
\end{equation}
and the Markovian limit is recovered.

For finite $\tau$ the underdamped  susceptibilities display oscillations, as shown in figure~\ref{Fig1}.
For memory kernels that are always positive, this feature is intimately related to the presence of a finite mass. In fact, as we will see in the next subsection, for overdamped dynamics oscillations appear only if the memory kernel has some negative parts. This behaviour of the susceptibilities is of course reflected in all quantities considered in the previous sections, as one can see from figure~\ref{Fig2}(a), for a system starting from an equilibrium condition, even if the dragging protocol $\lambda(t)=v t$ is linear. 
In the stationary state, memory effects are not visible anymore in the averages of position, work and entropy production rate (they grow linearly, see figure~\ref{Fig2}(b)) but oscillations are still present in the variance of work, which we have shown to follow the same formula for transient dynamics and for the stationary state. The non-monotonicity with time of the work variance is clearly due to the memory stored by the complex fluid along with inertial effects.
The variance of position and velocity are not shown in the figure as they are constant in both cases.
\begin{figure}[tb]
\includegraphics[width=0.95\textwidth, angle=0]{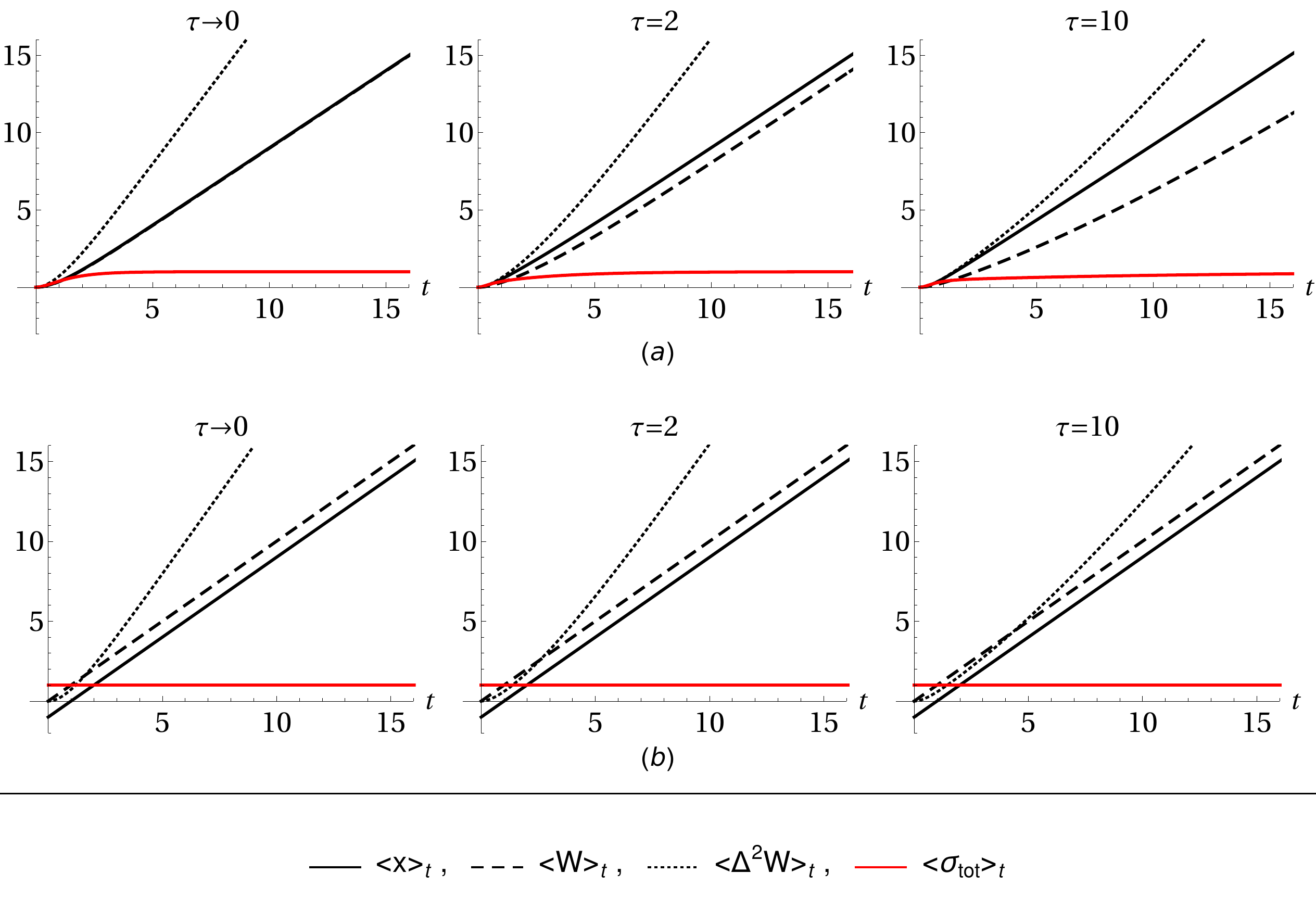}
\caption{For the overdamped case, evolution of the same quantities considered for the underdamped case starting (a) from equilibrium and (b) from a stationary state, for linear dragging protocol $\lambda(t)=vt$. Parameters are set as $\SC=1$, $\gamma_0=0.5$, $\gamma=0.5$ and $v=1$. For (a) we see that the main differences between the plots are visible for average work and its variance while for (b) this only happens for $\langle \Delta W \rangle^{ss}_{t}$ (that is, as we have shown in the previous sections, equal to the one starting from equilibrium $\langle \Delta W \rangle^{eq}_{t}$).}
\label{Fig5}
\end{figure}

Finally, if we consider an intrinsically oscillating driving protocol of the form $\lambda(t)=A \sin(\omega t)$, the effects of memory may determine an increase of the amplitude of the already present oscillations, both from equilibrium (figure~\ref{Fig3}(a)), and in the steady state (figure~\ref{Fig3}(b)). Panels on the left in figure~\ref{Fig3} represent the Markovian limit while panels on the right show an example for an exponential memory kernel with $\tau=2$. In the latter case,
the average position fluctuates more, and the entropy production rate can become negative (while having a positive average over one cycle in the steady oscillatory regime). 

We finish this section by noting that, even if the considered kernel is exponential, i.e. rapidly decaying, the effect of memory can extend to times much longer then the characteristic time $\tau$ of the kernel, as it can be seen from the figures.

\subsubsection{Overdamped dynamics}

\begin{figure}[tb]
\includegraphics[width=0.68\textwidth, angle=0]{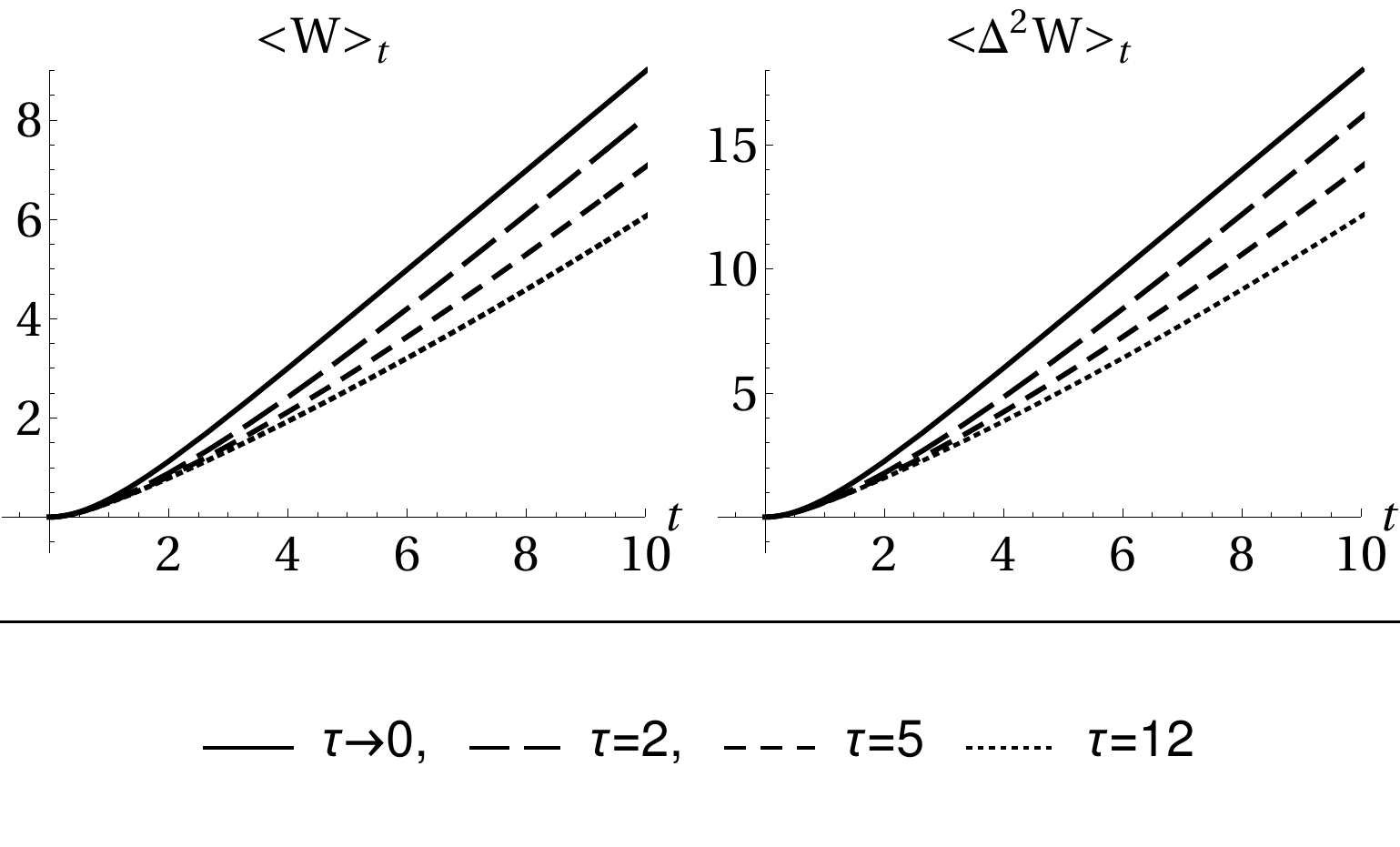}
\caption{Evolution of average work starting from equilibrium on the left panel and variance of work (equal from equilibrium or from stationary state) on the right, for the overdamped case (parameters as in the previous figure). }
\label{Fig6}
\end{figure}

\begin{figure}[tb]
\includegraphics[width=0.65\textwidth, angle=0]{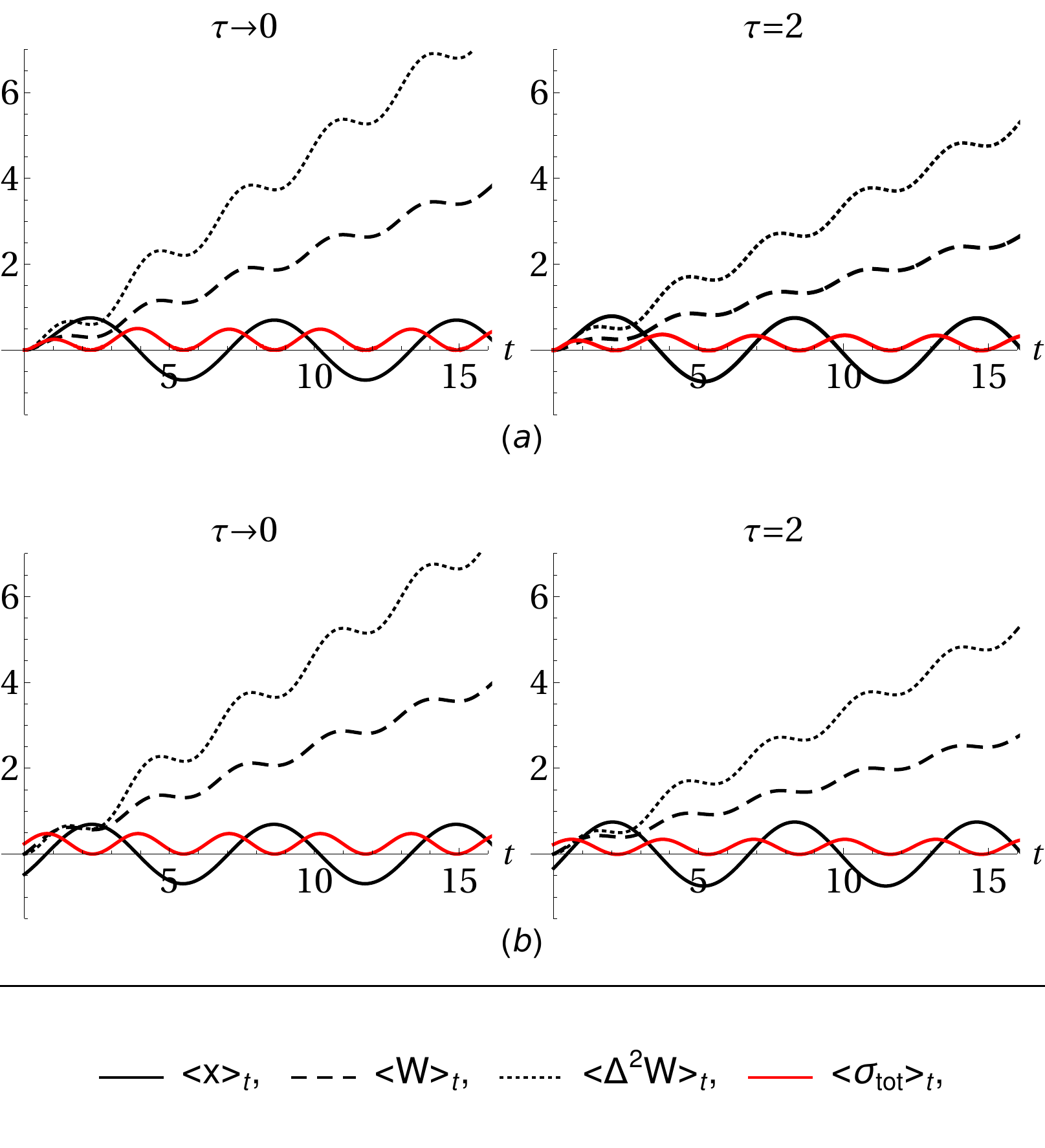}
\caption{For the overdamped case, time evolution of the already discussed thermodynamic quantities from equilibrium (a) and for $t_m \to - \infty$ (b) for intrinsically oscillating dragging protocol $\lambda(t)= A sin(\omega t)$. We chose $\SC=1$, $\gamma_0=0.5$, $\gamma=0.5$, $A=1$ and $\omega=1$. As before, we note important differences between the two columns concerning integrated quantities such as average work and variance, while average position and total entropy production rate are basically unaffected by the presence of memory.}
\label{Fig7}
\end{figure}

Here we consider the overdamped dynamics (\ref{GLE_over}) with the memory kernel
\begin{equation}\label{KernelExpOver2}
   \Gamma^{\text{exp}}(t) = 2 \gamma_{0} \delta(t)+
   \frac{\gamma}{\tau} \exp[-t/\tau]\quad\textrm{for } t\ge 0 
\end{equation}
The Dirac delta part is necessary in the overdamped limit for reasons of mathematical consistency, as shown in Appendix \ref{sec:limits} and \cite{nascimento2019non}.
Examples of susceptibilities for this kind of dynamics are displayed in figure~\ref{Fig4}. In particular, one sees that for $\gamma \geq 0$ the susceptibilities exhibit no oscillations, differently from the case with $\gamma < 0$ that is more alike to the underdamped case. The similarities between the overdamped GLE with negative memory kernels and underdamped dynamics has already been discussed in \cite{berner2018oscillating}. For this reason, in the following discussion we will mainly focus on the case with positive memory kernel.  

Figure~\ref{Fig5} shows the behaviour of the same quantities considered in the previous subsection for a linear dragging protocol $\lambda(t)=vt$. The differences between the plots for different values of $\tau$ are smaller than in the underdamped case shown in figure~\ref{Fig2}. This is due to the absence of oscillations. Nevertheless, for integrated quantities such as average work and its variance, the effect of a finite $\tau$ is evident for every $t>0$ (figure~\ref{Fig6}). In this case, the effects of memory determine a delay in the accumulation of thermodynamic work and in its variance. As a consequence, after some multiples of the characteristic time $\tau$, we observe a constant difference between the averages (starting from equilibrium) and variances (both from equilibrium and stationary state) of work for different values of $\tau$. This difference does not vanish in time and is also found for the entropy production (not shown). Thus, the exponential memory kernel influences the value of integrated quantities beyond its time scale $\tau$ even in the overdamped limit.

A similar behaviour is observed for the case of an intrinsically oscillating driving protocol $\lambda(t) = A \sin (\omega t)$. Indeed, figure~\ref{Fig7} shows that the effects of memory are again very evident for average work and variance, while average position and entropy production rate are not strongly affected.

\section{Conclusions}
The Gaussian process with memory is a classic in statistical mechanics. Yet, we have shown that further results can be derived for this process realised by a generalised Langevin equation for a particle driven by a harmonic strap with constant strength in a complex fluid. An explicit solution of the GLE is based on computing susceptibilities. In terms of these important dynamical quantities, several other expressions are derived. 

For generic protocols and initial Gaussian conditions, the quantities we computed for every time $t\ge 0$ are the average particle position \eqref{avgx}, its autocorrelation function \eqref{C(t',t'')} and hence its variance \eqref{varx}, the average work done on the system \eqref{avgwork}, its variance \eqref{VarWork}, and the entropy production rate \eqref{avgEntrTot}. These formulas can be simplified in some standard scenarios, e.g. starting from equilibrium or in steady states. Moreover, the variance of the work starting from equilibrium is equal to that for a steady state in a generalised sense and is proportional to the average of work starting from the same initial conditions. Since we can deal with various dragging protocols, this means that the two cumulants for the work \eqref{genVZC} generalise formulas by van Zon and Cohen~\cite{VanZon_cohen_Work}. 

Especially aiming at dealing with steady states, everything starts by introducing a new Laplace transform with arbitrary initial time $t_m$. The explicit dependence of the solution on $t_m$ along with the well-defined behaviour of  the susceptibilities for the limits $t \to 0$ and $t \to \infty$ allow us to recognise a steady state for a linear dragging protocol $\lambda(t)=vt$ as $t_m \to -\infty$. More in general, for an arbitrary protocol, this limit leads to a loss of the information about the initial state. We can interpret it as a generalised steady state. 

Going into some more details about the quantities calculated throughout the paper, for a steady state generated by a linear dragging protocol we recognise the same structure of the average of position and of velocity, and of their covariance matrix, as for usual Markov dynamics.  Finally, we are able to write the entropy production rate in terms of a quantity that we termed the retarded velocity, matching the usual velocity if no memory effects are included in the kernel. 

In conclusion, we note that this framework yields average quantities but also their variances. Hence it is used \cite{Memory_TUR} to derive one of the first examples of thermodynamic uncertainty relation~\cite{bar15,gin16,mae17,pie17,hor17,dec18,dec20,dit19} for systems with memory~\cite{vu19,vu20} .

\appendix

\section{Appendix: Limits of susceptibilities}\label{sec:limits}

In this section we discuss the limits of the position susceptibility defined in Laplace space as
\begin{equation}
\hat{\chi}_{x}(k) = [m k^{2}+k\hat{\Gamma}(k)+\SC]^{-1}
\end{equation}
along with the limit of its integral and and of its derivative, 
\begin{align} 
  \chi(t) \equiv \int_{0}^{t} \mathrm{d}t' \chi_{x}(t') \, , && \chi_{v} (t) \equiv \partial_{t} \chi_{x}(t) \, .
\end{align}
To this end we use that for a given function $g(t)$ it holds
\begin{align}
  \displaystyle\lim_{t\to 0} g(t) = \mathcal{L}^{-1}_{t} \left[\displaystyle\lim_{k\to \infty} \hat{g}(k) \right] \,,&& \displaystyle\lim_{t\to \infty} g(t) = \mathcal{L}^{-1}_{t} \left[\displaystyle\lim_{k\to 0} \hat{g}(k) \right] \, .
\end{align}
We first consider the long time limit of the susceptibilities 
\begin{equation} \label{lim1}
\begin{split}
  \displaystyle\lim_{t\to \infty} \chi_{x}(t) & = \mathcal{L}^{-1}_{t} \left[\displaystyle\lim_{k\to 0} \frac{1}{m k^{2}+k\hat{\Gamma}(k)+\SC} \right] \approx \mathcal{L}^{-1}_{t} \left[ \frac{1}{\SC} \right] =  \frac{2 \delta(t)}{\SC} \stackrel{t \to \infty}{=} 0 \, , \\
  \displaystyle\lim_{t\to \infty} \chi(t) & = \mathcal{L}^{-1}_{t} \left[\displaystyle\lim_{k\to 0} \frac{1}{k(m k^{2}+k\hat{\Gamma}(k)+\SC)} \right] \approx \mathcal{L}^{-1}_{t} \left[ \frac{1}{k \SC} \right] = \frac{\theta(t)}{\SC} \stackrel{t \to \infty}{=} 1/\SC \, ,\\
  \displaystyle\lim_{t\to \infty} \chi_{v}(t) & = 0 \, ,
\end{split}
\end{equation} 
where the last line immediately follows from the first line.
Note that all this limits do not depend on $m$ and hence they hold for both underdamped and overdamped dynamics. Things become different in the limit of $t\to 0$, where the the $m k^2$ term becomes dominant. Indeed, for underdamped dynamics, i.e. for finite $m$, we get
\begin{equation}
  \displaystyle\lim_{t\to 0} \chi^{\text{under}}_{x}(t) = \mathcal{L}^{-1}_{t} \left[\displaystyle\lim_{k\to \infty} [m k^{2}+k\hat{\Gamma}(k)+\SC]^{-1} \right] \approx \mathcal{L}^{-1}_{t} \left[ \frac{1}{m k^{2}} \right] = \frac{t}{m} \stackrel{t \to 0}{=} 0 \, ,
\end{equation}
where we used that $\displaystyle\lim_{k\to \infty} \frac{m k^2}{k\hat{\Gamma}(k)} \gg 1 $. In fact $\hat{\Gamma}(k) \stackrel{k \to \infty}{\propto} k$ would correspond to ballistic motion which we do not consider, see \cite{ViscGLE_Goy} for more details. As for its integral and derivative of course we have that
\begin{align}
  \displaystyle\lim_{t\to 0} \chi^{\textrm{under}}(t) = \displaystyle\lim_{t\to 0} \int_{0}^{t}\mathrm{d}t'\chi^{\text{under}}_{x}(t') \approx \frac{t^{2}}{2m} \stackrel{t \to 0}{=} 0 \,,&& \displaystyle\lim_{t\to 0} \chi^{\text{under}}_{v}(t) = \displaystyle\lim_{t\to 0} \partial_{t}\chi^{\text{under}}_{x}(t) \approx \frac{1}{m} \, .
\end{align}
We see that this result does not depend on the kernel form, in fact inertial effects dominate the particle behaviour in the small time limit. Moreover, it is clear from the last formulae that one can not simply take the massless limit $m \to 0 $ a posteriori to recover overdamped dynamics, because otherwise the limit of the susceptibilities would be ill defined. Instead, the correct procedure would correspond to calculate all the susceptibilities taking the mass $m$ exactly equal to zero a priori, i.e. one should compute
\begin{equation}
    \displaystyle\lim_{t\to 0} \chi^{\text{over}}_{x}(t) = \mathcal{L}^{-1}_{t} \left[\displaystyle\lim_{k\to \infty} [k\hat{\Gamma}(k)+\SC]^{-1} \right]
\end{equation}
that now depends on the details of the memory kernel. Consider for example a memory kernel consisting of a piece proportional to a Dirac delta, which alone would make the dynamics Markovian, plus a sum of exponentials.  
\begin{align}
  \Gamma^{\exp} (t) = 2 \gamma_{0}\delta(t)+ \sum_{i}\frac{\gamma_{i}}{\tau_{i}}\e^{-t/\tau_{i}} \,,
\end{align}
Its Laplace transform is equal to
\begin{align} 
  \hat{\Gamma}^{\exp} (k) = \gamma_{0}+ \sum_{i}\frac{\gamma_{i}}{1+k\tau_{i}} \,.
\end{align}
This is an important example, as a finite sum of appropriately chosen exponentials can approximate, up to a certain time scale, every memory kernel even if $\fric$ does not converge, see \cite{ViscGLE_Goy} for more details. 

Going back to the overdamped susceptibility, we have that
\begin{equation}
\begin{split}
    \displaystyle\lim_{t\to 0} \chi^{\text{exp, over}}_{x}(t) &= \mathcal{L}^{-1}_{t} \left[\displaystyle\lim_{k\to \infty} [k\hat{\Gamma}^{\mathrm{exp}}(k)+\SC]^{-1} \right] \approx \mathcal{L}^{-1}_{t} \left[\frac{1}{k \gamma_0(1 +\frac{1}{k \gamma_0}\sum_{i}\frac{\gamma_{i}}{\tau_{i}}+\frac{\SC}{k \gamma_0})}  \right] \approx \\
    & \approx \mathcal{L}^{-1}_{t} \left[\frac{1}{k \gamma_0} - \frac{1}{(k \gamma_0)^2}\left(\sum_{i}\frac{\gamma_{i}}{\tau_{i}}+\SC\right)\right] = \frac{1}{\gamma_0} - \frac{t}{ \gamma_0^2}\left(\sum_{i}\frac{\gamma_{i}}{\tau_{i}}+\SC\right)\,,
\end{split}
\end{equation}
\begin{equation}
  \displaystyle\lim_{t\to 0} \chi^{\text{exp, over}}(t) = \displaystyle\lim_{t\to 0} \int_{0}^{t}\mathrm{d}t'\chi_{x}^{\text{exp, over}}(t') \approx \frac{t}{\gamma_{0}} \stackrel{t \to 0}{=} 0 \,,
\end{equation}
\begin{equation}
    \displaystyle\lim_{t\to 0} \chi^{\text{exp, over}}_{v}(t) = \displaystyle\lim_{t\to 0} \partial_{t}\chi^{\text{exp, over}}_{x}(t) \approx - \frac{1}{ \gamma_0^2}\left(\sum_{i}\frac{\gamma_{i}}{\tau_{i}}+\SC\right)\, .
\end{equation}
We see that that, for this particular kernel, the overdamped limit requires the presence of the piece proportional to the Dirac delta. A more detailed discussion of this problem can be found in \cite{nascimento2019non}.

\section{Appendix: Calculation of $\mathcal{C} (t',t'')$}\label{sec:C(t',t')}
This appendix is dedicated to the calculation of the following quantity
\begin{equation}\label{Var3b}
\begin{split}
  &\mathcal{C} (t',t'')= \langle \phi (t')\phi (t'') \rangle = \int_{t_m}^{t'}\mathrm{d}s'\int_{t_m}^{t''}\mathrm{d}s'' \chi_{x} (t'-s') \chi_{x} \left( t''-s''\right) \langle \eta(s') \eta \left( s'' \right) \rangle =\\
  & \hspace{2cm}= k_{B}T\int_{t_m}^{t'}\mathrm{d}s'\int_{t_m}^{t''}\mathrm{d}s'' \chi_{x} (t'-s') \chi_{x} \left( t''-s''\right) \Gamma \left( |s' -s''| \right) \, .
  \end{split}
\end{equation}
In the last line we used the second fluctuation-dissipation theorem $\langle \eta (t') \eta \left( t'' \right) \rangle = k_{B} T \Gamma \left( |t' -t''| \right)$ that relates the correlation of the noise to the memory kernel. Taking the double modified Laplace transform of both sides of equation \eqref{Var3b} we get 
\begin{equation}\label{Var4}
\begin{split}
  &\hspace{4cm}\beta \mathcal{L}^{t_{m}}_{k'}\left[\mathcal{L}^{t_{m}}_{k''}\left[\mathcal{C} (t',t'')\right]\right] =\\
  &\hspace{0.4cm}=\int_{t_m}^{\infty}\mathrm{d}t'\e^{-k' t'} \int_{t_m}^{\infty}\mathrm{d}t''\e^{-k'' t''} \int_{t_m}^{t'}\mathrm{d}s'\int_{t_m}^{t''}\mathrm{d}s'' \chi_{x} (t'-s') \chi_{x} \left( t''-s''\right) \Gamma \left( |s' -s''| \right)= \\
  &\hspace{0.3cm}= \int_{t_m}^{\infty}\mathrm{d}s'\int_{t_m}^{\infty}\mathrm{d}s''\int_{s'}^{\infty}\mathrm{d}t'\e^{-k' t'} \int_{s''}^{\infty}\mathrm{d}t''\e^{-k'' t''} \chi_{x} (t'-s') \chi_{x} \left( t''-s''\right) \Gamma \left( |s' -s''| \right)=\\
  &= \int_{t_m}^{\infty}\mathrm{d}s'\e^{-k' s'}\int_{t_m}^{\infty}\mathrm{d}s''\e^{-k'' s''}\int_{0}^{\infty}\mathrm{d}u'\e^{-k' u'} \int_{0}^{\infty}\mathrm{d}u''\e^{-k'' u''} \chi_{x} (u') \chi_{x} \left( u''\right) \Gamma \left( |s' -s''| \right)=\\
  &\hspace{2cm} = \hat{\chi}_{x} (k') \hat{\chi}_{x} \left( k''\right)\int_{t_m}^{\infty}\mathrm{d}s'\e^{-k' s'}\int_{t_m}^{\infty}\mathrm{d}s''\e^{-k'' s''} \Gamma \left( |s' -s''| \right) \, ,
\end{split}
\end{equation}
where $\beta = 1/k_{B}T$ as usual. Moreover, we again used that $\int_{t_{m}}^{\infty} \mathrm{d}t \int_{t_m}^{t}\mathrm{d}t' = \int_{t_{m}}^{\infty}\mathrm{d}t' \int_{t'}^{\infty}\mathrm{d}t$ between the second and the third line and then we made the change of variable $u = t-s$. Focusing on the remaining integrals, we have that
\begin{equation}\label{Var5}
\begin{split}
  &\hspace{1cm}\int^{\infty}_{t_m}\mathrm{d}s'\e^{-k' s'}\int_{t_m}^{\infty}\mathrm{d}s''\e^{-k'' s''} \Gamma \left( |s' -s''| \right) =\\ &=\int_{t_m}^{\infty}\mathrm{d}s'\int_{t_m}^{\infty}\mathrm{d}s''\e^{-k' (s'-s'')}\e^{- s''(k'+k'')} \Gamma \left( |s' -s''| \right)=\\
  &\hspace{0.2cm}\stackrel{\sigma=s'-s''}{=} \int_{t_m}^{\infty}\mathrm{d}s''\e^{- s''(k'+k'')}\int_{t_m-s''}^{\infty}\mathrm{d\sigma}\e^{-k' \sigma} \Gamma \left( |\sigma| \right)= \\
  = \int_{t_m}^{\infty}&\mathrm{d}s''\e^{- s''(k'+k'')}\left(\int_{0}^{\infty}\mathrm{d\sigma}\e^{-k' \sigma} \Gamma \left( \sigma \right)+\int_{t_m-s''}^{0}\mathrm{d\sigma}\e^{-k' \sigma} \Gamma \left( -\sigma \right) \right)=\\
  = &\frac{\e^{- t_m(k'+k'')}}{k'+k''}\hat{\Gamma} (k')+\int_{t_m}^{\infty}\mathrm{d}s''\e^{- s''(k'+k'')}\int_{t_m-s''}^{0}\mathrm{d\sigma}\e^{-k' \sigma} \Gamma \left( -\sigma \right) \, ,
\end{split}
\end{equation}
 where in the last line we recognised the Laplace transform of $\Gamma(t)$ and used that
\begin{equation}
   \int_{t_m}^{\infty}\mathrm{d}s''\e^{- s''(k'+k'')}=\frac{\e^{- t_m(k'+k'')}}{k'+k''} \, .
\end{equation}
 As for the second term in the last line of equation \eqref{Var5}, using integration by parts we get
\begin{equation}
\begin{split}
&\hspace{2.5cm} \int_{t_m}^{\infty}\mathrm{d}s''\e^{- s''(k'+k'')}\int_{t_m-s''}^{0}\mathrm{d\sigma}\e^{-k' \sigma} \Gamma \left( -\sigma \right) =\\
&=-\left(\frac{\e^{- s''(k'+k'')}}{k'+k''}\int_{t_m-s''}^{0}\mathrm{d\sigma}\e^{-k' \sigma} \Gamma \left( -\sigma \right)\right)\Big|^{\infty}_{t_m} +\int_{t_m}^{\infty}\mathrm{d}s''\frac{\e^{- k''s''-k' t_m}}{k'+k''} \Gamma \left( s''-t_m \right) =\\
& \hspace{1.7cm}= \int_{t_m}^{\infty}\mathrm{d}s''\frac{\e^{- k'' s'' -k' t_m}}{k'+k''} \Gamma \left( s''-t_m \right)\stackrel{u=s''-t_m}{=}\frac{\e^{- t_m(k'+k'')}}{k'+k''}\hat{\Gamma} (k'') \, .
\end{split}
\end{equation}
where we noted that the first term in the second line is equal to zero. Going back to equation \eqref{Var5} and remembering that we started from \eqref{Var4} we finally obtain
\begin{equation}
  \beta \mathcal{L}^{t_{m}}_{k'}\left[\mathcal{L}^{t_{m}}_{k''}\left[\mathcal{C} (t',t'')\right]\right] =\hat{\chi}_{x} (k') \hat{\chi}_{x} \left( k''\right)\frac{\hat{\Gamma} (k')+\hat{\Gamma} (k'')}{k'+k''}\e^{- t_m(k'+k'')} \, .
\end{equation}
Recalling the definition of the position susceptibility via its Laplace transform and its relation with the memory kernel $\hat{\chi}_{x}(k) = [m k^{2}+k\hat{\Gamma}(k)+\SC]^{-1}$ and doing some algebra it is possible to show that
\begin{equation}
\begin{split}
  \beta \mathcal{L}^{t_{m}}_{k'}\left[\mathcal{L}^{t_{m}}_{k''}\left[\mathcal{C} (t',t'')\right]\right] =&\bigg[ \frac{\hat{\chi}_{x} (k')}{k''(k'+k'')} + \frac{\hat{\chi}_{x} (k'')}{k'(k'+k'')}+\\
  &-\SC \frac{\hat{\chi}_{x} (k')}{k'}\frac{\hat{\chi}_{x} (k'')}{k''}-m\hat{\chi}_{x} (k')\hat{\chi}_{x} (k'')\bigg]\e^{- t_m(k'+k'')} \, .
\end{split}
\end{equation}
The inverse transformation back to real time yields
\begin{equation}\label{Var10}
\begin{split}
\hspace{1.5cm}\beta \mathcal{C} (t',t'') =& \mathcal{L}^{t_{m},-1}_{t'}\left[\hat{\chi}_{x} (k')\e^{- t_m k'} \mathcal{L}^{t_{m},-1}_{t''}\left[\frac{\e^{- t_m k''}}{k''(k'+k'')}\right]\right] +\\ 
&+ \mathcal{L}^{t_{m},-1}_{t''}\left[\hat{\chi}_{x} (k'')\e^{- t_m k''} \mathcal{L}^{t_{m},-1}_{t'}\left[\frac{\e^{- t_m k'}}{k'(k'+k'')}\right]\right]+\\
&-\SC \mathcal{L}^{t_{m},-1}_{t'}\left[\frac{\hat{\chi}_{x} (k')\e^{- t_m k'}}{k'}\right]\mathcal{L}^{t_{m},-1}_{t''}\left[\frac{\hat{\chi}_{x} (k'')\e^{- t_m k''}}{k''}\right]+\\
&-m\mathcal{L}^{t_{m},-1}_{t'}\left[\hat{\chi}_{x} (k')\e^{- t_m k'}\right]\mathcal{L}^{t_{m},-1}_{t''}\left[\hat{\chi}_{x} (k'')\e^{- t_m k''}\right]\, .
\end{split}
\end{equation}
Using that 
\begin{align}
  \mathcal{L}^{t_{m},-1}_{t'}\left[\frac{1}{k'(k'+k'')}\right]= \frac{1}{k''}-\frac{\e^{- t' k''}}{k''} \,,&& \mathcal{L}^{t_{m},-1}_{t'}\left[\e^{- t_m k'}\right]= 2 \delta (t'-t_m)\, ,
\end{align}
along with the generalised convolution theorem, we are able to show that \eqref{Var10} becomes
\begin{equation}
\begin{split}
  \mathcal{C} (t',t'') =& k_{B}T \Big[\chi (t'-t_m)+\chi(t''-t_m)-\theta(t'-t'')\chi(t'-t'') +\\
  &-\theta(t''-t')\chi(t''-t')- \SC\chi (t'-t_m)\chi(t''-t_m) - m\chi_{x} (t'-t_m)\chi_{x}(t''-t_m) \Big]\, .
  \end{split}
\end{equation}



\end{document}